\documentclass[a4paper,11pt]{article}
\usepackage[a4paper,left=2.73cm,right=2.7cm,top=3cm,bottom=3.5cm]{geometry}

\usepackage{amsfonts,amsmath,amssymb,tabstackengine}
\usepackage{tikz}
\usepackage{array}
\usepackage{upgreek}
\usepackage{tensor}

\usepackage{comment}

\definecolor{Gray}{gray}{0.95}

\usepackage[colorlinks=true,linktocpage=true,linkcolor=blue,citecolor=blue]{hyperref}
\usepackage{graphicx}

\addtocontents{toc}{\protect\setcounter{tocdepth}{3}}
\numberwithin{equation}{section}

\def\cI{{\cal I}}

\def\cR{{\cal R}}

 \usepackage [latin1]{inputenc}

\begin{document}

\begin{titlepage}

\thispagestyle{empty}

\begin{center}

{\LARGE \textbf{Blackening S-folds}}

\vspace{40pt}
		
{\large \bf Adolfo Guarino}$\,^{a, b}$ \,  ,  \, {\large \bf Anik Rudra}$\,^c$ \,  ,  \, {\large \bf Colin Sterckx}$\,^d$  \,  \large{and} \, {\large \bf Mario Trigiante}$\,^{e}$

\vspace{25pt}
		
$^a$\,{\normalsize
Departamento de F\'isica, Universidad de Oviedo,\\
Avda. Federico Garc\'ia Lorca 18, 33007 Oviedo, Spain.}
\\[7mm]

$^b$\,{\normalsize
Instituto Universitario de Ciencias y Tecnolog\'ias Espaciales de Asturias (ICTEA), \\
Calle de la Independencia 13, 33004 Oviedo, Spain.}
\\[7mm]

$^c$\,{\normalsize
School of Physics and Mandelstam Institute for Theoretical Physics,\\ 
University of the Witwatersrand, Wits, 2050, South Africa.}
\\[7mm]

$^d$\,{\normalsize
INFN, Sezione di Padova,\\
Via Marzolo 8, 35131 Padova, Italy.}
\\[7mm]

$^e$\,{\normalsize
Dipartimento di Fisica, Politecnico di Torino, Corso Duca degli Abruzzi 24,\\
I-10129 Turin, Italy and INFN, Sezione di Torino, Italy.}
\\[10mm]

\texttt{adolfo.guarino@uniovi.es} \,\, , \,\, 
\texttt{anikrudra23@gmail.com} 
\texttt{colin.sterckx@pd.infn.it}  \,\, , \,\, \texttt{mario.trigiante@polito.it}

\vspace{20pt}

\vspace{20pt}
				
\abstract{
\noindent 
We construct the universal AdS$_{4}$ black hole that asymptotes to the $(\varphi,\chi)$-family of type IIB S-fold backgrounds dual to the conformal manifold of $\mathcal{N}=2$ S-fold CFT's. We present the explicit type IIB embedding of such a universal black hole for two particular asymptotics: the $\,\mathcal{N}=2\,$ S-fold with $\,\textrm{U}(2)\,$ symmetry at $\,(\varphi,\chi)=(0,0)\,$ and the $\,\mathcal{N}=4\,$ S-fold with $\,\textrm{SO}(4)\,$ symmetry at $\,(\varphi,\chi)=(1,0)$. As a byproduct, we also present a novel  $1/16$-BPS two-parameter family of $\,\textrm{AdS}_{2} \times \textrm{M}_{8}\,$ S-fold backgrounds with $\,\textrm{M}_{8}=\mathbb{H}^{2} \times \textrm{S}^{5}  \times \textrm{S}^{1}$ that features a parametrically-controlled scale separation.
}

\end{center}

\end{titlepage}

\tableofcontents

\hrulefill
\vspace{10pt}

\section{Introduction and motivation}

The study of four-dimensional (4D) supersymmetric black holes (BHs) and their embeddings in string theory and M-theory provides an exciting road to address fundamental questions in quantum gravity. Within the context of ten-dimensional (10D) and eleven-dimensional (11D) supergravities -- the low-energy limits of string theory and M-theory, respectively -- a particularly simple example of black hole, the so-called \textit{universal} black hole \cite{Romans:1991nq,Caldarelli:1998hg}, has captured much of attention in recent years. Its simplicity as a 4D solution, together with its various embeddings in string/M-theory, have made the universal black hole a central object of study also in light of the AdS/CFT correspondence \cite{Maldacena:1997re}.

Let us first introduce the universal black hole that will play a central role in this work. It is a black hole of Reissner--Nordstr\"om type, extremal (zero temperature) and, importantly, it asymptotes to an AdS$_{4}$ geometry with radius $L_{\textrm{AdS}_{4}}$ at infinity ($r \rightarrow \infty$). The 4D spacetime metric is given by
\begin{equation}
\label{metric_universal_BH_intro}
\begin{array}{rcl}
d s^2 = - f(r) \, d t^2 +  f(r)^{-1} \, d r^2 + r^{2} \, ds^{2}_{\mathbb{H}^{2}} \ ,
\end{array}
\end{equation}
with
\begin{equation}
\label{metric_universal_BH_2_intro}
f(r) = \left( \frac{r}{L_{\textrm{AdS}_{4}}} -\frac{L_{\textrm{AdS}_{4}}}{2\,r} \right)^2 \ ,
\end{equation}
and where $ds^{2}_{\mathbb{H}^{2}}$ is the metric on a Riemann surface $\Sigma_{\mathfrak{g}}$ of genus $\mathfrak{g}>1$ describing the horizon at radial distance $r_{H}^2 = \frac{1}{2} \,L^2_{\textrm{AdS}_{4}}$. The geometry (\ref{metric_universal_BH_intro})-(\ref{metric_universal_BH_2_intro}) is supported by a $\textrm{U}(1)$ Maxwell field $\,\mathcal{A}\,$ with field strength 
\begin{equation}
\label{flux_universal_BH_intro}
\mathcal{H} = d\mathcal{A} = p \, \textrm{vol}_{\mathbb{H}^{2}} \ , 
\end{equation}
endowing the BH with a constant magnetic charge $\,p\,$ being set by supersymmetry. In the near-horizon region $r \rightarrow r_{H}$, the geometry becomes $\,\textrm{AdS}_{2} \times \mathbb{H}^{2}\,$ with fixed ratios 
\begin{equation}
L^2_{\textrm{AdS}_{2}}=\frac{1}{2} \, L^2_{\mathbb{H}^{2}}=\frac{1}{4} \, L^2_{\textrm{AdS}_{4}} \ ,   
\end{equation}
between the relevant length scales. Finally, the Bekenstein--Hawking entropy of the universal AdS$_{4}$ BH is given by\footnote{$\text{Area}(\Sigma_{\mathfrak{g}})=L^2_{\Sigma_{\mathfrak{g}}} 4 \pi (\mathfrak{g}-1)\,$ where $\,L_{\Sigma_{\mathfrak{g}}}\,$ is the radius of a Riemann surface of genus $\mathfrak{g}>1$.}
\begin{equation}
\label{entropy_universal_intro}
s = \frac{\text{Area}(\Sigma_{\mathfrak{g}})}{4} = L^2_{\textrm{AdS}_{4}}\frac{(\mathfrak{g}-1)\pi}{2} \ .
\end{equation}

The universal AdS$_{4}$ BH we have just introduced can be framed within the context of minimal $\,\mathcal{N}=2$, $D=4\,$ supergravity with the graviphoton $\mathcal{A}$ being associated with the abelian gauge group $\textrm{G}_{\mathcal{N}=2}=\textrm{U}(1)$. In its simplest realisation, it describes a (\mbox{$1/4-$BPS}) supersymmetric solution of pure $\,\mathcal{N}=2\,$ supergravity in presence of a cosmological constant $\,\Lambda<0$. The bosonic part of the Lagrangian reads
\begin{equation}
\label{L_N=2_minimal_intro}
\mathcal{L}_{\mathcal{N}=2} = \left( \frac{R}{2} - \Lambda \right) *  1 - \tfrac{1}{2}  \, \mathcal{H} \wedge * \, \mathcal{H} \ ,
\end{equation}
where supersymmetry fixes the radius $L_{\textrm{AdS}_{4}}$ of the asymptotic AdS$_{4}$ region and the mass $\mu$ of the single (complex) gravitino in the theory as $L^{2}_{\textrm{AdS}_{4}} = |\mu|^{-2} = -3/\Lambda$. Via the AdS/CFT correspondence, the universal AdS$_{4}$ black hole is dual to a universal renormalisation group (RG) flow across dimensions \cite{Bobev:2017uzs}. This is a supersymmetric flow connecting a three-dimensional SCFT$_{3}$ dual to the AdS$_{4}$ vacuum at $r\rightarrow \infty$ (UV) to a one-dimensional superconformal quantum mechanics dual to the AdS$_{2}$ factor of the near-horizon geometry at $r \rightarrow r_{H}$ (IR). The RG flow is triggered by the action of a topological twist in the $\,\mathcal{N}=2\,$ SCFT$_{3}$ along its exact $\textrm{U}(1)_{\textrm{R}}$ superconformal R-symmetry. The precise identification of the SCFT$_{3}$ depends on the string/M-theory embedding of the universal AdS$_{4}$ black hole \cite{Azzurli:2017kxo}. Placing the SCFT$_{3}$ on $\textrm{S}^{1} \times \Sigma_{\mathfrak{g}}$ and computing the topologically twisted index of \cite{Benini:2015noa} at large $N$ (which counts the number of supersymmetric ground states), the Bekenstein--Hawking entropy (\ref{entropy_universal_intro}) is expected to be recovered in any string/M-theory embedding of the universal AdS$_{4}$ BH.

The above story actually generalises to a much larger class of quarter-BPS AdS$_{4}$ black holes in $\,\mathcal{N}=2\,$ supergravity coupled to matter fieds. This was first established in the impressive work \cite{Benini:2015eyy} (and its extension \cite{Benini:2016rke}) in the context of M-theory. The chosen $\,\mathcal{N}=2\,$ supergravity model was the so-called STU-model \cite{Cvetic:1999xp}. This model describes a consistent truncation of the maximal $\textrm{SO}(8)$ gauged supergravity \cite{deWit:1982ig} that arises from the dimensional reduction of 11D supergravity on the seven-sphere $\textrm{S}^{7}$ \cite{deWit:1986iy}. Unlike for the minimal $\,\mathcal{N}=2\,$ setup discussed before, the STU-model couples the $\,\mathcal{N}=2\,$ supergravity multiplet to $\,n_{v}=3\,$ abelian vector multiplets and $\,n_{h}=0\,$ hypermultiplets, \textit{i.e.} $\,(n_{v},n_{h})=(3,0)$. The three vector multiplets add three abelian vectors $\,\mathcal{A}^{i}\,$ and three complex scalars $\,z^{i}\,$ to the (bosonic) Lagrangian of the minimal model (\ref{L_N=2_minimal_intro}). Also the cosmological constant $\,\Lambda\,$ in (\ref{L_N=2_minimal_intro}) gets replaced by a non-trivial scalar potential $\,V(z^{i},\,\bar{z}^{\bar{\imath}})$. The STU-model has a larger $\textrm{G}_{\mathcal{N}=2}=\textrm{U}(1)^{4}\subset \textrm{SO}(8)$ gauge group which allows for a generalisation of the universal AdS$_{4}$ BH in (\ref{metric_universal_BH_intro})-(\ref{flux_universal_BH_intro}) to include three additional (magnetic \cite{Benini:2015eyy} and electric \cite{Benini:2016rke}) charges associated with the vectors in the $n_{v}=3$ vector multiplets, as well as non-trivial profiles $z^{i}(r)$ for the scalars \cite{Cacciatori:2009iz}. Lastly, the AdS$_{4}$ vacuum chosen in \cite{Benini:2015eyy,Benini:2016rke} to describe the asymptotic region ($r \rightarrow \infty$) of the black holes was the maximally supersymmetric\footnote{$\mathcal{N}=2$ within the STU-model.} $\textrm{AdS}_{4} \times \textrm{S}^{7}$ Freund--Rubin background of 11D supergravity \cite{Freund:1980xh} AdS/CFT dual to the ABJM \cite{Aharony:2008ug} SCFT$_{3}$. Then the explicit computations carried out in \cite{Benini:2015eyy,Benini:2016rke} using localisation techniques showed a perfect matching between the topologically twisted index of ABJM on $\,\textrm{S}^{1} \times \Sigma_{\mathfrak{g}}\,$ at large $N$ and the gravitational entropy of this multi-charge class of AdS$_{4}$ black holes with running scalars in the STU-model. Finally, the universal AdS$_{4}$ BH in (\ref{metric_universal_BH_intro})-(\ref{flux_universal_BH_intro}) is recovered as the simplest solution with (constant) vanishing scalars $\,z^{1,\,2,\,3}=i\,$ and non-zero magnetic charge along the exact $\textrm{U}(1)_{\textrm{R}}\subset \textrm{U}(1)^{4}$ superconformal R-symmetry of the ABJM theory.

A holographic counting of black hole microstates has also been performed in the context of massive type IIA strings \cite{Azzurli:2017kxo,Hosseini:2017fjo,Benini:2017oxt}. The relevant $\,\mathcal{N}=2\,$ supergravity model is the extended STU-model of  \cite{Guarino:2017pkw}. This model describes a $\textrm{U}(1)^{2}$-invariant sector of the maximal $\textrm{ISO}(7)$ supergravity \cite{Guarino:2015qaa} that arises from the dimensional reduction of massive type IIA supergravity on the six-sphere $\textrm{S}^{6}$ \cite{Guarino:2015vca}. The extended STU-model couples the $\,\mathcal{N}=2\,$ supergravity multiplet to $\,n_{v}=3\,$ abelian vector multiplets, as in the previous M-theory case, but also to $\,n_{h}=1\,$ hypermultiplet, \textit{i.e.} $\,(n_{v},n_{h})=(3,1)$. The need to include the (universal) hypermultiplet stems from the fact that, upon dimensional reduction to four dimensions, the ten-dimensional dilaton lies in such a hypermultiplet. The abelian gauge group of the extended STU-model in the massive IIA context turns out to be $\,\textrm{G}_{\mathcal{N}=2}=\textrm{U}(1)^{3} \times \mathbb{R}\,$ featuring a non-compact generator. The universal AdS$_{4}$ black hole was presented in \cite{Guarino:2017eag}. Its asymptotic ($r \rightarrow \infty$) region approaches the $\mathcal{N}=2$ $\textrm{AdS}_{4} \times \textrm{S}^{6}$ background of massive IIA supergravity with $\textrm{U}(3)$ symmetry dual to a super-Chern--Simons-matter theory at level $\,k\,$ (given by the Romans mass parameter \cite{Romans:1985tz}) and simple gauge group $\textrm{SU}(N)$ \cite{Guarino:2015jca}. Constructing explicitly the non-universal multi-charge black holes with running scalars becomes much more complicated in the presence of hypermultiplets. However, although such BH solutions have not been constructed yet, a careful analysis of the horizon data was enough to carry out a holographic counting of BH microstates in \cite{Azzurli:2017kxo,Hosseini:2017fjo,Benini:2017oxt} along the lines of the M-theory case.

In this work we continue the above program and present, amongst other solutions, the universal AdS$_{4}$ black hole in the context of type IIB strings. 
The paper is summarised as follows. In Section~\ref{sec:STU-model} we construct the extended STU-model of relevance in the type IIB context. We obtain it as a $\mathbb{Z}_{2}\times\mathbb{Z}_{2}$-invariant sector of the maximal $[\textrm{SO}(6) \times \textrm{SO}(1,1)]  \ltimes \mathbb{R}^{12}$ supergravity that arises from the reduction of type IIB supergravity on $\textrm{S}^{5} \times \textrm{S}^{1}$ including an $\,\textrm{SO}(1,1)\,$ duality twist along the $\,\textrm{S}^{1}$ \cite{Inverso:2016eet}. We reformulate the model as an $\,\mathcal{N}=2\,$ supergravity coupled to $(n_{v},n_{h})=(3,4)$ matter multiplets with an abelian gauge group $\,\textrm{G}_{\mathcal{N}=2}=\textrm{U}(1)^{2} \times \mathbb{R}^{2}$. In Section~\ref{sec:solutions} we first carry out an exhaustive classification of $\,\textrm{AdS}_{2} \times \Sigma_{\mathfrak{g}}\,$ solutions suitable to describe the near-horizon geometry of black holes. Then we present the universal AdS$_{4}$ black hole that asymptotes \textit{any} solution in the two-parameter $(\varphi,\chi)$-family of $\,\mathcal{N}=2\,$ $\,\textrm{AdS}_{4} \times \textrm{S}^{1} \times \textrm{S}^{5}\,$ S-fold backgrounds of type IIB supergravity dual to the conformal manifold of $\,\mathcal{N}=2\,$ S-fold CFT$_{3}$'s \cite{Bobev:2021yya}. Black hole solutions asymptoting the family of $\,\mathcal{N}=2\,$ ${\rm AdS}_4$ S-folds have been built within a truncation described by a  $D=4$ Einstein-Maxwell theory in \cite{Bobev:2023bxs}. We extend these results by directly constructing the universal BHs as solutions to the BPS equations within a consistent $\mathcal{N}=2$ truncation. In Section~\ref{sec:IIB_uplift}, fetching techniques from the $\textrm{E}_{7(7)}$ Exceptional Field Theory (ExFT) of \cite{Hohm:2013uia}, we present the uplift of the universal AdS$_{4}$ black holes that asymptote the $\mathcal{N}=2$ S-fold with $\textrm{U}(2)$ symmetry at $\,(\varphi,\chi)=(0,0)\,$ \cite{Guarino:2020gfe}, as well as the $\mathcal{N}=4$ S-fold with $\textrm{SO}(4)$ symmetry at $\,(\varphi,\chi)=(1,0)\,$ \cite{Inverso:2016eet}\footnote{See \cite{Giambrone:2021wsm} for a judicious rewriting of the $\mathcal{N}=4\,\&\,\textrm{SO}(4)$ S-fold of \cite{Inverso:2016eet} closer to the nomenclature used in this work.}. As a byproduct, we also discuss some higher-dimensional aspects of an unexpected two-parameter family of $1/4$-BPS $\,\textrm{AdS}_{2} \times \mathbb{H}^{2}\,$ solutions that uplift to $1/16$-BPS $\,\textrm{AdS}_{2} \times \textrm{M}_{8}\,$ S-fold backgrounds of type IIB supergravity, with $\textrm{M}_{8}=\mathbb{H}^{2} \times \textrm{S}^{5}  \times \textrm{S}^{1}$, and feature parametrically-controlled scale separation. In Section~\ref{sec:conclusions} we conclude and discuss some future directions. Two appendices are included at the end with technical aspects of the supergravity construction as well as with a set of first-order BPS equations in matter-coupled $\,\mathcal{N}=2\,$ supergravity.

\section{An STU-model from type IIB on \texorpdfstring{$\textrm{S}^{1} \times \textrm{S}^{5}$}{S1 x S5}}
\label{sec:STU-model}

Our starting point is the maximal ($\mathcal{N}=8$) supergravity in four dimensions (4D) with gauge group
\begin{equation}
\textrm{G}=[\textrm{SO}(6) \times \textrm{SO}(1,1)]  \ltimes \mathbb{R}^{12} \subset \textrm{SL}(8) \subset \textrm{E}_{7(7)} \ .
\end{equation}
Labeling the fundamental representation of $\textrm{SL}(8)$ by an index $\mathsf{A}=1,\ldots,8$, the bosonic sector of the theory consists of the metric (\mbox{spin-$2$}) field $g_{\mu\nu}$, $28$ electric $\mathcal{A}_{\mu}{}^{[\mathsf{AB}]}$ and $28$ magnetic $\tilde{\mathcal{A}}_{\mu}{}_{[\mathsf{AB}]}$ vector (\mbox{spin-$1$}) fields, and $70$ \mbox{spin-$0$} fields serving as coordinates in the scalar geometry described by the coset space
\begin{equation}
\mathcal{M}_{\textrm{scal}}=\frac{\textrm{E}_{7(7)}}{\textrm{SU}(8)} \ . 
\end{equation}
As stated in the introduction, this maximal supergravity has been shown to accommodate a rich structure of $\textrm{AdS}_{4}$ solutions \cite{DallAgata:2011aa,Guarino:2019oct,Guarino:2020gfe,Guarino:2021hrc} that uplift to non-geometric $\textrm{AdS}_{4} \times \textrm{S}^{1} \times \textrm{S}^{5}$ S-fold backgrounds of type IIB supergravity \cite{Inverso:2016eet,Guarino:2019oct,Bobev:2019jbi,Guarino:2020gfe,Bobev:2020fon,Giambrone:2021zvp,Giambrone:2021wsm} (see \cite{Guarino:2022tlw} for an overview).

All the couplings in the Lagrangian of maximal supergravity in 4D are encoded in an object called the \textit{embedding tensor} \cite{deWit:2007mt}. Such a tensor specifies how the (local) gauge group $\textrm{G}$ of the supergravity, also known as the \textit{gauging}, is embedded into the (global) $\textrm{E}_{7(7)}$ duality group of the ungauged theory. More concretely, the embedding tensor $X_{\mathcal{MN}}{}^{\mathcal{P}}$ carries fundamental indices $\mathcal{M}=1,\ldots,56$ of $\textrm{E}_{7(7)}$ and is subject to a set of linear constraints so that $X_{\mathcal{MN}}{}^{\mathcal{P}}$ lives in the $\bf{912}$ irreducible representation (irrep) of $\textrm{E}_{7(7)}$. In addition, in order to define a consistent gauged supergravity, the embedding tensor must obey a set of quadratic constraints (all the details can be found in \cite{deWit:2007mt}, see also \cite{Samtleben:2008pe,Trigiante:2016mnt} for reviews). We frame this work within the $\textrm{G}=[\textrm{SO}(6) \times \textrm{SO}(1,1)] \ltimes \mathbb{R}^{12}$ gauged maximal supergravity with a gauging of the dyonic type investigated in \cite{Inverso:2016eet}. Using the group-theoretical branching rule $\,\bf{56} \rightarrow \bf{28} + \bf{28}'\,$ under $\textrm{E}_{7(7)} \supset \textrm{SL}(8)$, which translates into an index splitting of the form $T_{\mathcal{M}}=\left( T_{[\mathsf{AB}]}\,,\,T^{[\mathsf{AB}]} \right)$, the embedding tensor of the theory has components
\begin{equation}
\label{X_components_STU}
\begin{array}{llllll}
{X_{[\mathsf{AB}] [\mathsf{CD}]}}^{[\mathsf{EF}]} &=& - X_{[\mathsf{AB}] \phantom{[\mathsf{EF}]} [\mathsf{CD}]}^{\phantom{[\mathsf{AB}]} [\mathsf{EF}]} &=& -8 \, \delta_{[\mathsf{A}}^{[\mathsf{E}} \, \eta_{\mathsf{B}][\mathsf{C}} \, \delta_{\mathsf{D}]}^{\mathsf{F}]}  &  , \\[4mm]
X^{[\mathsf{AB}] \phantom{[\mathsf{CD}]}[\mathsf{EF}]}_{\phantom{[\mathsf{AB}]}[\mathsf{CD}]} &=& - {X^{[\mathsf{AB}] [\mathsf{EF}]}}_{[\mathsf{CD}]} &=& -8 \, \delta_{[\mathsf{C}}^{[\mathsf{A}} \, \tilde{\eta}^{\mathsf{B}][\mathsf{E}} \, \delta_{\mathsf{D}]}^{\mathsf{F}]} &  ,
\end{array}
\end{equation}
given in terms of the two symmetric matrices
\begin{equation}
\eta_{\mathsf{AB}} = g \, \textrm{diag} (\, \mathbb{I}_{5} \, , \, 0  \, , \, 0 \, ,  \, 1 \,)
\hspace{10mm} \textrm{ and } \hspace{10mm}
\tilde{\eta}^{\mathsf{AB}} = g\, c \,\,  \textrm{diag} (\, 0_{5}  \, , \,  -1   \, ,  \, 1 \, ,  \, 0\,) \ .
\end{equation}
The constants $\,g\,$ and $\,c\,$ denote the gauge coupling and an electromagnetic parameter in the maximal supergravity, respectively \cite{Inverso:2016eet}.

\subsection{\texorpdfstring{$\mathbb{Z}_{2} \times \mathbb{Z}_{2}$}{Z2 x Z2} invariant sector: STU model}

Up to our knowledge, all the known solutions in the $[\textrm{SO}(6) \times \textrm{SO}(1,1)] \ltimes \mathbb{R}^{12}$ maximal supergravity (including the flows of \cite{Guarino:2021kyp,Guarino:2022tkh}) were constructed within an Einstein-scalar setup, namely, vector fields were always set to zero. In this work we will be interested in charged black holes, so both electric and magnetic vector fields will generically be turned on. In order to have a simple supergravity model where to search for charged solutions we will mod-out the $[\textrm{SO}(6) \times \textrm{SO}(1,1)] \ltimes \mathbb{R}^{12} $ maximal supergravity by a specific $\mathbb{Z}_2 \times \mathbb{Z}_{2}$ group. As we show in Appendix~\ref{app:N=2_models}, this particular $\,\mathbb{Z}_2 \times \mathbb{Z}_{2}\,$ gives rise to the most general $\,\mathcal{N}=2\,$ supergravity model capturing the entire $(\varphi,\chi)$-family of AdS$_{4}$ vacua of the maximal theory as $\,\mathcal{N}=2\,$ solutions. In the $\textrm{SL}(8)$ basis, the $\mathbb{Z}_2 \times \mathbb{Z}_{2}$ group is generated by
\begin{equation}
\label{Z2xZ2_action}
\begin{array}{rcl}
\mathcal{O}_1=\textrm{diag}{(1,-1,-1,-1,-1,1,1,1)} & , \\[2mm]
\mathcal{O}_2=\textrm{diag}{(-1,1,-1,1,-1,1,-1,1)} & ,
\end{array}
\end{equation}
together with the identity element $\mathbb{I}$ and $\mathcal{O}_{1} \mathcal{O}_{2}$. Therefore, we will construct the $\mathbb{Z}_2 \times \mathbb{Z}_{2}$ invariant sector of the $[\textrm{SO}(6) \times \textrm{SO}(1,1)] \ltimes \mathbb{R}^{12}$ maximal supergravity by retaining those fields that are invariant under the action of (\ref{Z2xZ2_action}).

The vectors $\,\mathcal{A}_{\mu}{}^{\mathcal{M}}=( \mathcal{A}_{\mu}{}^{[\mathsf{AB}]}, \tilde{\mathcal{A}}_{\mu}{}_{[\mathsf{AB}]} )\,$ in the maximal theory that are invariant under the action generated by (\ref{Z2xZ2_action}) are
\begin{equation}
\label{Z2xZ2_inv_vectors}
\mathcal{A}_{\mu}{}^{[24]}
\hspace{2mm},\hspace{2mm}
\mathcal{A}_{\mu}{}^{[35]}
\hspace{2mm},\hspace{2mm}
\mathcal{A}_{\mu}{}^{[17]}
\hspace{2mm},\hspace{2mm}
\mathcal{A}_{\mu}{}^{[68]}  \ ,
\end{equation}
together with their magnetic duals. They span the gauge group
\begin{equation}
\label{N=2_gauging}
\textrm{G}_{\mathcal{N}=2} \,\,=\,\, \textrm{U}(1)_{1} \times \textrm{U}(1)_{2} \times \mathbb{R}_{1} \times \mathbb{R}_{2}  \,\,\subset\,\, \textrm{G}  \ ,
\end{equation}
where each factor is respectively generated by the $\textrm{SL}(8)\subset \textrm{E}_{7(7)}$ generators\footnote{We use the conventions in the appendix of \cite{Guarino:2015tja} regarding the generators of $\textrm{E}_{7(7)}$ in the $\textrm{SL}(8)$ basis (see eqs $(60)$-$(61)$ therein). These split into the $63$ generators of $\textrm{SL}(8)\subset \textrm{E}_{7(7)}$, which are of the form $t_{\mathsf{A}}{}^{\mathsf{B}}$ with $t_{\mathsf{A}}{}^{\mathsf{A}}=0$, and the completion to $\textrm{E}_{7(7)}$ given by $70$ generators of the form $t_{\mathsf{ABCD}}=t_{[\mathsf{ABCD}]}$.}
\begin{equation}
({t_2}^4 - {t_4}^2)\,,\,({t_3}^5-{t_5}^3) \subset \mathfrak{so}(6)
\hspace{5mm} \textrm{ and } \hspace{5mm}
{t_7}^1\,,\,{t_6}^8 \subset \mathbb{R}^{12} \ .
\end{equation}
The scalar geometry of the $\,\mathbb{Z}_{2} \times \mathbb{Z}_{2}\,$ invariant sector we are interested in is identified with
\begin{equation}
\label{M_scal}
\mathcal{M}_\textrm{scal} = \left[\frac{\textrm{SU}(1,1)}{\textrm{U}(1)}\right]^3 \times  \frac{\textrm{SO}(4,4)}{\textrm{SO}(4) \times \textrm{SO}(4)} \subset \frac{\textrm{E}_{7(7)}}{\textrm{SU}(8)}\ ,
\end{equation}
and is encoded in the following coset representative. The $\,\left[\textrm{SU}(1,1)/\textrm{U}(1)\right]^3\,$ factor describes a special K\"ahler (SK) geometry, involves the following generators 
\begin{equation}
\begin{array}{cclccclc}
g_{\varphi_{1}} &=& t_{2}{}^{2} + t_{4}{}^{4} + t_{6}{}^{6} + t_{8}{}^{8} - t_{1}{}^{1} - t_{3}{}^{3} - t_{5}{}^{5} - t_{7}{}^{7} & ,  & g_{\chi_{1}} &=& t_{2468} & ,  \\[2mm]
g_{\varphi_{2}} &=& t_{7}{}^{7} + t_{1}{}^{1} + t_{6}{}^{6} + t_{8}{}^{8} - t_{2}{}^{2} - t_{3}{}^{3} - t_{4}{}^{4} - t_{5}{}^{5} & , & g_{\chi_{2}} &=& t_{7168} & , \\[2mm]
g_{\varphi_{3}} &=& t_{5}{}^{5} + t_{3}{}^{3} + t_{6}{}^{6} + t_{8}{}^{8} - t_{1}{}^{1} - t_{2}{}^{2} - t_{4}{}^{4} - t_{7}{}^{7} & , & g_{\chi_{3}} &=& t_{5368} & , 
\end{array}
\end{equation}
and has a coset representative
\begin{equation}
\label{eq:VSKN8}
\mathcal{V}_{\textrm{SK}} = e^{12 \, \sum \chi_{i}\, g_{\chi_{i}} } \, e^{-\frac{1}{4} \sum \varphi_{i} \, g_{\varphi_{i}}} \in  \left[\frac{\textrm{SU}(1,1)}{\textrm{U}(1)}\right]^3 \ ,
\end{equation}
with $i=1,2,3$. The $\,\textrm{SO}(4,4)/\left(\textrm{SO}(4) \times \textrm{SO}(4)\right)\,$ factor describes a quaternionic K\"ahler (QK) geometry, involves the following generators 
\begin{equation}
\begin{array}{cclccclc}
g_{\tilde{\varphi}_{1}} &=& t_{5}{}^{5} - t_{3}{}^{3}  & ,  & g_{\tilde{\chi}_{1}} &=& t_{3}{}^{5} & , \\[2mm]
g_{\tilde{\varphi}_{2}} &=& t_{8}{}^{8} - t_{6}{}^{6}  & ,  & g_{\tilde{\chi}_{2}} &=& t_{6}{}^{8} & , \\[2mm]
g_{\tilde{\varphi}_{3}} &=& t_{2}{}^{2} - t_{4}{}^{4} & ,  & g_{\tilde{\chi}_{3}} &=& t_{4}{}^{2}   & , 
\end{array}
\end{equation}
together with
\begin{equation}
\begin{array}{cclccclccclccclccclcl}
g_{\phi} &=& t_{1}{}^{1} - t_{7}{}^{7}  & , & g_{\zeta^{0}} &=&   t_{2578}   & , &  g_{\zeta^{1}} &=&  t_{2378} & , & g_{\zeta^{2}} &=&   t_{2576} & , &  g_{\zeta^{3}} &=&  t_{4578} & ,  \\[2mm]
g_{\sigma} &=& - t_{7}{}^{1} & , & g_{\tilde{\zeta}_{0}} &=&  t_{3467}   & , & g_{\tilde{\zeta}_{1}} &=& t_{4567}  & , & g_{\tilde{\zeta}_{2}} &=& t_{3478}  & , & g_{\tilde{\zeta}_{3}} &=& t_{2367}  & ,  
\end{array}
\end{equation}
and has a coset representative
\begin{equation}
\label{eq:VQKN8}
\mathcal{V}_{\textrm{QK}} = e^{12 \sum \left(  \zeta^{A} \, g_{\zeta^A}  +  \tilde{\zeta}_{A} \, g_{\tilde{\zeta}_{A}} \right) }  \, e^{\sigma  g_{\sigma}  +  \sum \tilde{\chi}_{a} g_{\tilde{\chi}_{a}} } \, e^{\phi \, g_{\phi} + \frac{1}{2} \sum \tilde{\varphi}_{a} g_{\tilde{\varphi}_{a}} } \in \frac{\textrm{SO}(4,4)}{\textrm{SO}(4) \times \textrm{SO}(4)} \ ,
\end{equation}
with $A=0,1,2,3$ and $a=1,2,3$. Finally, the coset representatives in (\ref{eq:VSKN8}) and (\ref{eq:VQKN8}) can be used to introduce a scalar-dependent matrix 
\begin{equation}
\label{M_MN_4D}
\mathcal{M}_{\mathcal{MN}} = \left(  \mathcal{V} \, \mathcal{V}^{t} \right)_{\mathcal{MN}} \in \textrm{E}_{7(7)} 
\hspace{8mm} \textrm{ with } \hspace{8mm}
\mathcal{V} = \mathcal{V}_{\textrm{SK}} \, \mathcal{V}_{\textrm{QK}} \in \frac{\textrm{E}_{7(7)}}{\textrm{SU}(8)} \ . 
\end{equation}

As we will show in a moment, this $\,\mathbb{Z}_{2} \times \mathbb{Z}_{2}\,$ invariant sector of the maximal theory can be reformulated as an $\,\mathcal{N}=2\,$ supergravity coupled to $\,n_{v}=3\,$ vector multiplets and $\,n_{h}=4\,$ hypermultiplets.

\subsection{\texorpdfstring{$\mathcal{N}=2$}{N=2} reformulation of the STU-model}

Let us present the reformulation of the $\,\mathbb{Z}_{2} \times \mathbb{Z}_{2}\,$ invariant model described above as an $\mathcal{N}=2$ supergravity coupled to $n_{v}=3$ vector multiplets and $n_{h}=4$ hypermultiplets. The $\mathcal{N}=2$ Lagrangian is given by
\begin{equation}
\label{L_N2_STU}
\begin{array}{lll}
L &=&   \left( \frac{R}{2} - V \right) *  1 - K_{i \bar{\jmath}} \,  dz^{i} \wedge * \, d\bar{z}^{\bar{\jmath}} - h_{uv} \, Dq^{u} \wedge * \, Dq^{v}   \\[2mm]
&+& \frac{1}{2} \, \cI_{\Lambda \Sigma} \,  \mathcal{H}^\Lambda \wedge * \, \mathcal{H}^\Sigma +  \frac{1}{2} \, \cR_{\Lambda \Sigma} \, \mathcal{H}^\Lambda \wedge \mathcal{H}^\Sigma + L_{\textrm{top}} \ .
\end{array}
\end{equation}

The scalar fields in the vector multiplets and hypermultiplets  -- we respectively denote them $\,z^{i}\,$ ($i=1,\ldots,n_{v}$) and $\,q^{u}\,$ ($u=1,\ldots, 4\,n_{h}$) -- span the factorised scalar geometry previously identified, namely,
\begin{equation}
\mathcal{M}_\textrm{scal} = \left[\frac{\textrm{SU}(1,1)}{\textrm{U}(1)}\right]^3 \times  \frac{\textrm{SO}(4,4)}{\textrm{SO}(4) \times \textrm{SO}(4)} \ .
\end{equation}
While the scalars $\,z^{i}\,$ in the vector multiplets are not charged under the gauge group (\ref{N=2_gauging}), the covariant derivatives of the scalars $\,q^{u}\,$ in the hypermultiplets are given by
\begin{equation}
\label{Dq}
Dq^{u} = \partial q^{u} - \mathcal{A}^{M} \, \Theta_{M}{}^{\alpha} \, k_{\alpha}{}^{u} \ ,
\end{equation}
with $\,M=1,\ldots,2(n_{v}+1)\,$ and $\,\alpha=1,\ldots,4\,$. These covariant derivatives depend on the embedding tensor $\,\Theta\,$ and the Killing vectors $\, k_{\alpha}{}^{u}\,$ specifying the gauging of $\textrm{G}_{\mathcal{N}=2}$ in (\ref{N=2_gauging}).

The two-form field strengths $\,\mathcal{H}^{\Lambda}\,$ for the (electric) vector fields $\,\mathcal{A}^{\Lambda}\,$ entering (\ref{L_N2_STU}), with $\Lambda=(0,i)$, read
\begin{equation}
\label{H_electric}
\mathcal{H}^{\Lambda} = d\mathcal{A}^{\Lambda} - \tfrac{1}{2} \Theta^{\Lambda \alpha}\mathcal{B}_{\alpha} \ ,
\end{equation}
and generically involve two-form tensor fields $\,\mathcal{B}_{\alpha}\,$ as dictated by the embedding tensor $\,\Theta$. The two-forms $\,\mathcal{B}_{\alpha}$, as well as the magnetic vectors $\tilde{\mathcal{A}}_{\Lambda}$, do not carry independent dynamics but enter the topological term $L_{\textrm{top}}$ in (\ref{L_N2_STU}). We have explicitly verified that the full $\mathcal{N}=2$ Lagrangian in (\ref{L_N2_STU}) matches the one computed directly in the maximal theory using the formulation of \cite{deWit:2007mt}.

\subsubsection{Vector multiplets}

We will parameterise the three complex scalars $\,z^{i}\,$ in the vector multiplets as
\begin{equation}
z^{i} = -\chi_{i} + i e^{-\varphi_{i}}  
\hspace{8mm} \textrm{ with } \hspace{8mm} i=1,2,3 \ . 
\end{equation}
They describe the special K\"ahler geometry $\,\mathcal{M}_\textrm{SK}=[\textrm{SU}(1,1)/\textrm{U}(1)]^3$ in terms of a set of holomorphic sections
\begin{equation}
X^{M}=\left(  X^{\Lambda}(z),F_{\Lambda}(z)\right)    
\hspace{8mm} \textrm{ with } \hspace{8mm}
\Lambda=(0,i) \ .
\end{equation}
The K\"ahler potential associated with $\mathcal{M}_\textrm{SK}$ is given by $K=-\log \left(  i \left\langle  X, \bar{X}\right\rangle\right)$ in terms of the symplectic product of vectors
\begin{equation}
\left\langle  U, W\right\rangle \equiv U^{M} \Omega_{MN}  V^{N} = U_{\Lambda} V^{\Lambda} - U^{\Lambda} V_{\Lambda}  \ ,
\end{equation}
defined using the $\,\Omega_{MN}\,$ antisymmetric invariant matrix of $\,\textrm{Sp}(8)\,$. We take the holomorphic sections to be
\begin{equation}
\left(  X^{0},X^{1},X^{2},X^{3}, F_{0},F_{1},F_{2},F_{3}\right) = \left(   - z^{1} z^{2} z^{3} , -z^{1} , -z^{2} , -z^{3} , 1 , z^{2} z^{3} , z^{3} z^{1} , z^{1} z^{2} \right) \ ,
\end{equation}
which satisfy $\,F_{\Lambda}=\partial\mathcal{F}/\partial X^{\Lambda}\,$ for a prepotential function
\begin{equation}
\mathcal{F} = - 2 \sqrt{X^{0} X^{1} X^{2} X^{3}} \ . 
\end{equation}
The K\"ahler potential gives rise to a K\"ahler metric of the form
\begin{equation}
ds^2_{\textrm{SK}} = K_{i\bar{\jmath}} \, dz^{i} \, d\bar{z}^{\bar{\jmath}} = \tfrac{1}{4} \displaystyle\sum_{a=1}^{3} \left(  d\varphi_{a}^2 + e^{2 \varphi_{a}} d\chi_{a}^2 \right) \ .
\end{equation}

The kinetic terms and generalised theta angles for the dynamical vectors in the Lagrangian (\ref{L_N2_STU}) are encoded in the matrix
\begin{equation}
\label{N_matrix_N2}
\mathcal{N}_{\Lambda \Sigma} = \bar{F}_{\Lambda \Sigma}+ 2 \, i \,  \frac{\textrm{Im}(F_{\Lambda \Gamma})X^{\Gamma}\,\,\textrm{Im}(F_{\Sigma \Delta})X^{\Delta}}{\textrm{Im}(F_{\Omega \Phi})X^{\Omega}X^{\Phi}}
\hspace{4mm} \textrm{ where } \hspace{4mm}
F_{\Lambda \Sigma}=\partial_{\Lambda}\partial_{\Sigma} \mathcal{F} \ .
\end{equation}
An explicit computation yields
\begin{equation}
\mathcal{N}_{\Lambda\Sigma}=
\frac{1}{n}
\left(
\begin{array}{cccc}
 -i e^{\varphi_{1}+\varphi_{2}+\varphi_{3}} & n_{1} & n_{2} & n_{3}\\
n_{1} & -i
   e^{\varphi_{1}-\varphi_{2}-\varphi_{3}} \, c_{2} \, c_{3} & 
   n_{12} & n_{13}  \\
n_{2} &
n_{12} & -i e^{-\varphi_{1}+\varphi_{2}-\varphi_{3}} \, c_{1} \, c_{3} & 
n_{23}\\
n_{3}  & n_{13} & n_{23} & -i e^{-\varphi_{1}-\varphi_{2}+\varphi_{3}} \, c_{1} \, c_{2}
\end{array}
\right) \ ,
\end{equation}
in terms of $\,c_{i} \equiv ( 1+ e^{2 \varphi_{i}} \, \chi_{i}^2)\,$ and where we have introduced the quantities
\begin{equation}
\begin{array}{llll}
n & \equiv & \Big( 1+ \displaystyle\sum_{k} e^{2 \varphi_{k}}  \chi_{k}^{2}  \Big)  + 2 \, i \, e^{\varphi_{1}+\varphi_{2}+\varphi_{3}}\,  \chi_{1} \, \chi_{2} \, \chi_{3} \ , &  \\[2mm]
n_{i} & \equiv & e^{2 \varphi_{i}} \chi_{i}+i \, e^{\varphi_{1}+\varphi_{2}+\varphi_{3}} \chi_{j} \, \chi_{k} & \hspace{10mm} (i \neq j \neq k) \ ,\\[2mm]
n_{ij} & \equiv & e^{-\varphi_{k}} \, c_{k} \,  (e^{\varphi_{k}} \,  \chi_{k}+ i \,  e^{\varphi_{i} + \varphi_{j}} \, \chi_{i} \chi_{j} )  & \hspace{10mm} (i \neq j \neq k) \ .
\end{array}
\end{equation}
Defining $\,\mathcal{R}_{\Lambda \Sigma}\equiv \textrm{Re}(\mathcal{N}_{\Lambda \Sigma})\,$ and $\,\mathcal{I}_{\Lambda \Sigma}\equiv \textrm{Im}(\mathcal{N}_{\Lambda \Sigma})\,$, we can introduce a symmetric, real and negative-definite scalar matrix
\begin{equation}
\label{M_scalar_matrix_N2_STU}
\mathcal{M}(z^{i}) = \left( 
\begin{array}{cc}
\mathcal{I} + \mathcal{R} \mathcal{I}^{-1} \mathcal{R}   & -\mathcal{R} \mathcal{I}^{-1} \\
- \mathcal{I}^{-1} \mathcal{R} & \mathcal{I}^{-1}
\end{array}
\right) \ ,
\end{equation}
which will be used later on when describing the attractor equations in (\ref{attractor_eqs_STU}). This matrix satisfies the relations $\,{\mathcal{M}_{MN} \mathcal{V}^{N}=i \Omega_{MN}\mathcal{V}^{N}}\,$ and $\,\mathcal{M}_{MN} D_{z}\mathcal{V}^{N}=-i \Omega_{MN}D_{z}\mathcal{V}^{N}$, where $\,\mathcal{V}^{M} \equiv e^{K/2} \, X^{M}\,$ is a redefined (non-holomorphic) set of sections with K\"ahler covariant derivatives given by $\,D_{z}\mathcal{V}^{M} = \partial_{z} \mathcal{V}^{M} + \frac{1}{2} (\partial_{z} K) \mathcal{V}^{M}$.

\subsubsection{Hypermultiplets}

The four hypermultiplets contain sixteen real scalars serving as coordinates in the quaternionic K\"ahler geometry $\mathcal{M}_\textrm{QK}=\textrm{SO}(4,4)/\textrm{SO}(4)\times \textrm{SO}(4)$. The geometry is given by
\begin{equation}
\label{ds_QK_STU}
\begin{array}{lll}
ds_{\textrm{QK}}^2 &=&  h_{u v} \, dq^{u} dq^{v} \\[2mm]
&=& \widetilde{K}_{a \bar{b}} \, d\tilde{z}^{a} \, d\bar{\tilde{z}}^{\bar{b}}
+  \, d \phi \,  d \phi - \frac{1}{4} \, e^{2\phi} \,  ( d\vec{\zeta}\,)^{T}  \widetilde{\mathcal{M}}_{8} \, d \vec{\zeta} \\[2mm]
	&+& \frac{1}{4} e^{4 \phi} \left[ d \sigma +\frac{1}{2}\, (\vec{\zeta}\,)^{T} \, \mathbb{C} \, d \vec{\zeta} \,  \right]  \,   \left[ d \sigma +\frac{1}{2}\, (\vec{\zeta}\,)^{T} \, \mathbb{C} \, d \vec{\zeta} \, \right]  \ ,
\end{array}
\end{equation}
with $\,\mathbb{C}=-\Omega\,$ and where we have introduced the notation $\,\vec{\zeta}\equiv(\zeta^{A},\tilde{\zeta}_{A})$, with $\,A=0,1,2,3$, to describe the scalars parameterising the Heisenberg fiber of the QK geometry.

The matrix $\,\widetilde{\mathcal{M}}_{8} \,$ depends on the complex scalars  $\,\tilde{z}^{a}=\tilde{\chi}_{a}+ie^{-\tilde{\varphi}_{a}}\,$, with $\,a = 1,2,3$, parameterising the special K\"ahler manifold $\,\mathcal{M}_{\widetilde{\textrm{SK}}}\,$ of the c-map. More specifically, it has the structure given in (\ref{M_scalar_matrix_N2_STU}), namely,
\begin{equation}
\label{M_scalar_matrix_SK_STU}
\widetilde{\mathcal{M}}_{8} = \left( 
\begin{array}{cc}
\tilde{\mathcal{I}} + \tilde{\mathcal{R}} \mathcal{I}^{-1} \tilde{\mathcal{R}}   & -\tilde{\mathcal{R}} \tilde{\mathcal{I}}^{-1} \\
- \tilde{\mathcal{I}}^{-1} \tilde{\mathcal{R}} & \tilde{\mathcal{I}}^{-1}
\end{array}
\right) \ ,
\end{equation}
with the building blocks
\begin{equation}
\tilde{\mathcal{I}} =  -  e^{-\sum \tilde{\varphi}_{a}}
\left(
\begin{array}{cccc}
 1+ \sum e^{2 \tilde{\varphi }_a} \tilde{\chi }_a^2 & -e^{2 \tilde{\varphi }_1} \tilde{\chi }_1 & -e^{2 \tilde{\varphi }_2} \tilde{\chi }_2 & -e^{2 \tilde{\varphi }_3} \tilde{\chi }_3 \\[2mm]
 -e^{2 \tilde{\varphi }_1} \tilde{\chi }_1 & e^{2 \tilde{\varphi }_1} & 0 & 0 \\[2mm]
 -e^{2 \tilde{\varphi }_2} \tilde{\chi }_2 & 0 & e^{2 \tilde{\varphi }_2} & 0 \\[2mm]
 -e^{2 \tilde{\varphi }_3} \tilde{\chi }_3 & 0 & 0 & e^{2 \tilde{\varphi }_3}
\end{array}
\right) \ , 
\end{equation}
and
\begin{equation}
\tilde{\mathcal{R}} = 
\left(
\begin{array}{cccc}
 -2 \tilde{\chi }_1 \tilde{\chi }_2 \tilde{\chi }_3 & \tilde{\chi }_2 \tilde{\chi }_3 & \tilde{\chi }_1 \tilde{\chi }_3 & \tilde{\chi }_1 \tilde{\chi }_2 \\[2mm]
 \tilde{\chi }_2 \tilde{\chi }_3 & 0 & -\tilde{\chi }_3 & -\tilde{\chi }_2 \\[2mm]
 \tilde{\chi }_1 \tilde{\chi }_3 & -\tilde{\chi }_3 & 0 & -\tilde{\chi }_1 \\[2mm]
 \tilde{\chi }_1 \tilde{\chi }_2 & -\tilde{\chi }_2 & -\tilde{\chi }_1 & 0
\end{array}
\right) \ .
\end{equation}
As a result, the complex scalars $\,\tilde{z}^{a}\,$ span a scalar geometry $\,\mathcal{M}_{\widetilde{\textrm{SK}}}=\left[\textrm{SU}(1,1)/\textrm{U}(1)\right]^3 \subset \mathcal{M}_{\textrm{QK}}\,$ with metric
\begin{equation}
\label{metric_SK_c-map_STU}
ds^2_{\widetilde{\textrm{SK}}}=\widetilde{K}_{a \bar{b}} \, d\tilde{z}^{a} \, d\bar{\tilde{z}}^{\bar{b}} = \tfrac{1}{4} \sum_{a=1}^{3} \left(  d\tilde{\varphi}_{a}^2 + e^{2 \tilde{\varphi}_{a}} d\tilde{\chi}_{a}^2 \right) \ .
\end{equation}
The geometry $\,\mathcal{M}_{\widetilde{\textrm{SK}}}\,$ can be described in terms of a cubic prepotential $\,\tilde{\mathcal{F}}=-\tilde{Z}^{1} \tilde{Z}^{2} \tilde{Z}^{3}/\tilde{Z}^{0}\,$ \cite{deWit:1991nm}, a set of sections $\,\tilde{Z}=(\tilde{Z}^{A},\tilde{G}_{A})\,$ with $\tilde{Z}^{A}=\left(1,\tilde{z}^{1},\tilde{z}^{2},\tilde{z}^{3}\right)$ and $\tilde{G}_{A}=\partial \tilde{F}/\partial \tilde{Z}^{A}$, and a K\"ahler potential $\,\tilde{K}=-\log \left(  i \left\langle  \tilde{Z}, \bar{\tilde{Z}}\right\rangle\right)$ from which the metric (\ref{metric_SK_c-map_STU}) follows.

\subsubsection{Embedding tensor, scalar potential and topological term}

The STU-model incorporates a gauging of $\,\textrm{G}_{\mathcal{N}=2} =  \textrm{U}(1)_{1} \times \textrm{U}(1)_{2} \times \mathbb{R}_{1} \times \mathbb{R}_{2}\,$ specified by an embedding tensor $\,\Theta\,$ of the form
\begin{equation}
\label{Embedding_Tensor_STU}
\Theta_{M}{}^{\alpha} = g \, 
\left(
\begin{array}{cccc}
 0 & -1 & 0 & 0 \\
 0 & 0 & -1 & 0 \\
 1 & 0 & 0 & 0 \\
 0 & 0 & 0 & 1 \\
 \hline
 0 & c & 0 & 0 \\
 0 & 0 & 0 & 0 \\
 c & 0 & 0 & 0 \\
 0 & 0 & 0 & 0 \\
\end{array}
\right) \ ,
\end{equation}
and involves four Killing vectors $\,k_{\alpha}\,$ ($\alpha=1,2,3,4$) of the QK geometry (\ref{ds_QK_STU}). The non-compact $\,\mathbb{R}_{1}\,$ and $\,\mathbb{R}_{2}\,$ factors associated with the $\,k_{1}\,$ and $\,k_{2}\,$ isometries are dyonically gauged by the combinations of vectors $\,\mathcal{A}^{2}+c\,\tilde{\mathcal{A}}_{2}\,$ and $\,-\mathcal{A}^{0}+c\,\tilde{\mathcal{A}}_{0}\,$, respectively. The compact $\,\textrm{U}(1)_{1}\,$ factor in the gauge group associated with the $\,k_{3}\,$ isometry is electrically gauged by $\,-\mathcal{A}^{1}\,$. Lastly, the compact $\,\textrm{U}(1)_{2}\,$ factor in the gauge group associated with the $\,k_{4}\,$ isometry is electrically gauged by $\,\mathcal{A}^{3}\,$. The dictionary between these vectors and the ones of the maximal theory in (\ref{Z2xZ2_inv_vectors}) reads
\begin{equation}
\label{A^M_4D}
\mathcal{A}^{[24]} = -\mathcal{A}^{1}
\hspace{5mm} , \hspace{5mm} 
\mathcal{A}^{[35]} = \mathcal{A}^{3}
\hspace{5mm} , \hspace{5mm} 
\mathcal{A}^{[17]} = \mathcal{A}^{2}
\hspace{5mm} , \hspace{5mm} 
\mathcal{A}^{[68]} = -\mathcal{A}^{0} \ ,
\end{equation}
and similarly for the magnetic ones. From a ten-dimensional perspective, the two compact isometries descend from the $\textrm{SO}(6)$ isometries of the internal S$^5$ in the type IIB reduction. Lastly, the electric field-strengths (\ref{H_electric}) entering the Lagrangian (\ref{L_N2_STU})
\begin{equation}
\label{H_electric_STU}
\mathcal{H}^{0} = d\mathcal{A}^{0} - \tfrac{gc}{2} \, \mathcal{B}_{2} 
\hspace{5mm} , \hspace{5mm}
\mathcal{H}^{1} = d\mathcal{A}^{1}
\hspace{5mm} , \hspace{5mm}
\mathcal{H}^{2} = d\mathcal{A}^{2} - \tfrac{gc}{2} \, \mathcal{B}_{1} 
\hspace{5mm} , \hspace{5mm}
\mathcal{H}^{3} = d\mathcal{A}^{3}  \ ,
\end{equation}
where $\,\mathcal{B}_{1}\,$ and $\,\mathcal{B}_{2}\,$ are two-form potentials whose role will be clarified later on when discussing the topological term $\,L_{\textrm{top}}\,$ in (\ref{L_N2_STU}).

\subsubsection*{Scalar potential}

In order to compute the scalar potential $\,V\,$ in (\ref{L_N2_STU}) and the covariant derivatives of the hyperscalars in (\ref{Dq}), we must identify the four Killing vectors entering the gauging. The first Killing vector is given by
\begin{equation}
\label{k1_vec_STU}
k_{1} = \partial_{\sigma} \ ,
\end{equation}
and has a prepotential
\begin{equation}
P_{1}=\left(
\begin{array}{c}
 0 \\
 0 \\
 \frac{1}{2} e^{2 \phi } \\
\end{array}
\right) \ ,
\end{equation}
whereas the second, third and fourth Killing vectors are
\begin{equation}
\label{k2&k3&k4_vecs_STU}
\begin{array}{lcl}
k_{2} &=& \partial_{\tilde{\chi}_{2}} + \zeta^{0}\partial_{\zeta^{2}} - \tilde{\zeta}_{2} \partial_{\tilde{\zeta}_{0} } - \zeta^{3} \partial_{\tilde{\zeta}_{1} } - \zeta^{1} \partial_{\tilde{\zeta}_{3} }   \ , \\[4mm]
k_{3} &=& 2 \tilde{\chi}_{3} \partial_{\tilde{\varphi}_{3}} + \left(  e^{-2 \tilde{\varphi}_{3}} - \tilde{\chi}_{3}^2   -1 \right) \partial_{\tilde{\chi}_{3}   }  + \zeta^{3}\partial_{\zeta^{0}} - \tilde{\zeta}_{2} \partial_{\zeta^{1}} - \tilde{\zeta}_{1} \partial_{\zeta^{2} } - \zeta^{0}\partial_{\zeta^{3} }\\[2mm]
&+& \tilde{\zeta}_{3}\partial_{\tilde{\zeta}_{0}} + \zeta^{2} \partial_{\tilde{\zeta}_{1}} + \zeta^{1} \partial_{\tilde{\zeta}_{2}} - \tilde{\zeta}_{0} \partial_{\tilde{\zeta}_{3}} \ , \\[4mm]
k_{4} &=& -2 \tilde{\chi}_{1} \partial_{\tilde{\varphi}_{1}} - \left(  e^{-2 \tilde{\varphi}_{1}} - \tilde{\chi}_{1}^2 -1 \right) \partial_{\tilde{\chi}_{1}   }  - \zeta^{1}\partial_{\zeta^{0}} + \zeta^{0} \partial_{\zeta^{1}} + \tilde{\zeta}_{3} \partial_{\zeta^{2} } + \tilde{\zeta}_{2}\partial_{\zeta^{3} }\\[2mm]
&-& \tilde{\zeta}_{1}\partial_{\tilde{\zeta}_{0}} + \tilde{\zeta}_{0} \partial_{\tilde{\zeta}_{1}} - \zeta^{3} \partial_{\tilde{\zeta}_{2}} - \zeta^{2} \partial_{\tilde{\zeta}_{3}} \ .
\end{array}
\end{equation}
The three Killing vectors in (\ref{k2&k3&k4_vecs_STU}) can be expressed in a very compact form in terms of the sections $\tilde{Z}$ in the SK basis of the QK geometry. Following \cite{Erbin:2014hsa} they can be expressed as
\begin{equation}
\label{Killing_vectors_234}
\begin{array}{lll}
k_{2,3,4} &=& \left[ (\mathbb{U}_{2,3,4} \, \tilde{Z})^{A} \partial_{\tilde{Z}^{A}} + \textrm{c.c.}\right] + (\mathbb{U}_{2,3,4} \, \vec{\zeta} \, )^{T}  \, \partial_{\vec{\zeta}} \ ,
\end{array}
\end{equation}
in terms of the electric section components $\,\tilde{Z}^{A}\,$ and the three $\mathbb{U}$-matrices
\begin{equation}
\label{U_matrices_12}
\mathbb{U}_{2} =
\left(
\begin{array}{cccccccc}
 0 & 0 & 0 & 0 & 0 & 0 & 0 & 0 \\
 0 & 0 & 0 & 0 & 0 & 0 & 0 & 0 \\
 1 & 0 & 0 & 0 & 0 & 0 & 0 & 0 \\
 0 & 0 & 0 & 0 & 0 & 0 & 0 & 0 \\
 0 & 0 & 0 & 0 & 0 & 0 & -1 & 0 \\
 0 & 0 & 0 & -1 & 0 & 0 & 0 & 0 \\
 0 & 0 & 0 & 0 & 0 & 0 & 0 & 0 \\
 0 & -1 & 0 & 0 & 0 & 0 & 0 & 0 \\
\end{array}
\right)
\hspace{3mm} \textrm{ , } \hspace{3mm}
\mathbb{U}_{3} =
\left(
\begin{array}{cccccccc}
 0 & 0 & 0 & 1 & 0 & 0 & 0 & 0 \\
 0 & 0 & 0 & 0 & 0 & 0 & -1 & 0 \\
 0 & 0 & 0 & 0 & 0 & -1 & 0 & 0 \\
 -1 & 0 & 0 & 0 & 0 & 0 & 0 & 0 \\
 0 & 0 & 0 & 0 & 0 & 0 & 0 & 1 \\
 0 & 0 & 1 & 0 & 0 & 0 & 0 & 0 \\
 0 & 1 & 0 & 0 & 0 & 0 & 0 & 0 \\
 0 & 0 & 0 & 0 & -1 & 0 & 0 & 0 \\
\end{array}
\right) \ ,
\end{equation}
and
\begin{equation}
\label{U_matrix_4}
\mathbb{U}_{4} =
\left(
\begin{array}{cccccccc}
 0 & -1 & 0 & 0 & 0 & 0 & 0 & 0 \\
 1 & 0 & 0 & 0 & 0 & 0 & 0 & 0 \\
 0 & 0 & 0 & 0 & 0 & 0 & 0 & 1 \\
 0 & 0 & 0 & 0 & 0 & 0 & 1 & 0 \\
 0 & 0 & 0 & 0 & 0 & -1 & 0 & 0 \\
 0 & 0 & 0 & 0 & 1 & 0 & 0 & 0 \\
 0 & 0 & 0 & -1 & 0 & 0 & 0 & 0 \\
 0 & 0 & -1 & 0 & 0 & 0 & 0 & 0 \\
\end{array}
\right) \ .
\end{equation}
The associated prepotentials are also constructed directly from the $\mathbb{U}$-matrices in (\ref{U_matrices_12}) and (\ref{U_matrix_4}) as
\begin{equation}
P_{2,3,4}=\left(
\begin{array}{c}
- \sqrt{2} \,  e^{\frac{\widetilde{K}}{2}+\phi}  \, \textrm{Re} \left[  \tilde{Z}^{T} \mathbb{C} \, \mathbb{U}_{2,3,4} \, \vec{\zeta} \right]  \\[2mm]
- \sqrt{2} \,  e^{\frac{\widetilde{K}}{2}+\phi}  \, \textrm{Im} \left[  \tilde{Z}^{T} \mathbb{C} \, \mathbb{U}_{2,3,4} \, \vec{\zeta} \right]  \\[2mm]
-  \tfrac{1}{4} e^{2 \phi} (\vec{\zeta}\,)^{T} \mathbb{C} \, \mathbb{U}_{2,3,4} \, \vec{\zeta}  + e^{\widetilde{K}} \tilde{Z}^{T} \mathbb{C} \, \mathbb{U}_{2,3,4} \, \bar{\tilde{Z}}
\end{array}\right) \ .
\end{equation}
Equipped with the above data it is straightforward to construct the scalar potential in (\ref{L_N2_STU}) for the STU-model using the $\,\mathcal{N}=2\,$ symplectically covariant expression \cite{Michelson:1996pn,deWit:2005ub}
\begin{equation}
\label{VN2}
V_{\mathcal{N}=2} =  4 \,  \mathcal{V}^{M}  \, \bar{\mathcal{V}}^{N}    \, \mathcal{K}_{M}{}^{u}  \, h_{uv} \,  \mathcal{K}_{N}{}^{v}
+ \mathcal{P}^{x}_{M} \, \mathcal{P}^{x}_{N} \left( K^{i\bar{\jmath}} \, D_{i}\mathcal{V}^{M} \, D_{\bar{\jmath}} \bar{\mathcal{V}}^{N}  - 3 \, \mathcal{V}^{M} \, \bar{\mathcal{V}}^{N} \right) \ ,
\end{equation}
with $\,\mathcal{V}^{M} \equiv e^{K/2} \, X^{M}\,$ and $\,D_{i}\mathcal{V}^{M}=\partial_{z^{i}} \mathcal{V}^{M} + \frac{1}{2} (\partial_{z^{i}} K)\mathcal{V}^{M}$, and where we have introduced symplectic Killing vectors $\,\mathcal{K}_{M}{}^{u}  \equiv \Theta_{M}{}^{\alpha} \, k_{\alpha}{}^{u}\,$ and moment maps $\,\mathcal{P}_{M}^{x}  \equiv \Theta_{M}{}^{\alpha} \, P_{\alpha}^{x}\,$ in order to maintain symplectic covariance \cite{Klemm:2016wng}.

\subsubsection*{Topological term}

The topological term $\,L_{\textrm{top}}\,$ in (\ref{L_N2_STU}) reads
\begin{equation}
\begin{array}{rcl}
L_{\textrm{top}} &=& \frac{1}{2}\,\Theta^{\Lambda\alpha}\,\mathcal{B}_\alpha\wedge d\tilde{\mathcal{A}}_\Lambda+\frac{1}{8}\,\Theta^{\Lambda\alpha} \, \Theta_\Lambda{}^{\beta}\,\mathcal{B}_\alpha\wedge \mathcal{B}_\beta \\[2mm]
& = & \frac{1}{2} \, g \, c \left[ \mathcal{B}_{1} \wedge d\tilde{\mathcal{A}}_{2} + \mathcal{B}_{2} \wedge d\tilde{\mathcal{A}}_{0} +\frac{g}{4} \left(  \mathcal{B}_{1} \wedge \mathcal{B}_{1} - \mathcal{B}_{2} \wedge \mathcal{B}_{2}  \right) \right] \ ,
\end{array}    
\end{equation}
where the two-form potentials $\,\mathcal{B}_{1}\,$ and $\,\mathcal{B}_{2}\,$ (previously introduced in (\ref{H_electric_STU})) are associated with the two non-compact isometries $k_{1}$ and $k_{2}$ that are gauged dyonically by the embedding tensor (\ref{Embedding_Tensor_STU}). These two-form potentials do not carry an independent dynamics as they are dual to scalar currents by virtue of the equations of motion of the magnetic vectors. For example, the equation of motion of $\,\tilde{\mathcal{A}}_{2}\,$ yields the duality relation
\begin{equation}
\label{dB1_current}
d\mathcal{B}_{1}  \,\, \propto \,\,  e^{4 \phi}  * \left[ D \sigma +\frac{1}{2}\, (\vec{\zeta}\,)^{T} \, \mathbb{C} \, D \vec{\zeta} \,  \right]  \ .
\end{equation}
The equation of motion of $\,\tilde{\mathcal{A}}_{0}\,$ then sets a duality relation between $\,d\mathcal{B}_{2}\,$ and a different scalar current.

\section{Supersymmetric solutions}
\label{sec:solutions}

In this section we present some analytic and supersymmetric solutions of the $\,\mathcal{N}=2\,$ model without and with vector fields.

\subsection{\texorpdfstring{$\mathcal{N}=2$ AdS$_{4}$}{N=2 AdS4} solutions (S-folds)}
\label{sec:general_N=2_AdS4}

The vanishing of all fermionic supersymmetry variations translates into the conditions \cite{Louis:2012ux}
\begin{equation}
\label{N2_cond_AdS4_STU}
\begin{array}{rlll}
X^{M}\, \mathcal{K}_{M} &=& 0 & , \\[2mm]
\left[ \,  \partial_{z^{i}} X^{M} + (\partial_{z^{i}} K) X^{M} \, \right] \, \mathcal{P}^{x}_{M} & = & 0 & , \\[2mm]
S_{\mathcal{AB}} \, \epsilon^{\mathcal{B}} &=& \frac{1}{2} \, \mu \, \epsilon^{\ast}_{\mathcal{A}} & ,
\end{array}
\end{equation}
where $\,S_{\mathcal{AB}}=\frac{1}{2} \, e^{K/2} X^{M}  \mathcal{P}^{x}_{M} (\sigma^{x})_{\mathcal{AB}}\,$ is the gravitino mass matrix expressed in terms of Pauli matrices $\,(\sigma^{x})_{\mathcal{AB}}\,$ and $\,|\mu|=L^{-1}_{\textrm{AdS}_{4}}\,$. The three conditions in (\ref{N2_cond_AdS4_STU}) follow from the vanishing of the hyperini, gaugini and gravitini supersymmetry variations, respectively.

The algebraic system (\ref{N2_cond_AdS4_STU}) can be solved in full generality. It gives maximally symmetric AdS$_{4}$ solutions with radius 
\begin{equation}
\label{L2_AdS4}
L^{2}_{\textrm{AdS}_{4}}=-\frac{3}{V_{0}}=\frac{c}{g^2} \ ,   
\end{equation}
preserving $\mathcal{N}=2$ supersymmetry. The locus in the scalar field space is given by
\begin{equation}
\label{AdS4_VEV_STU_1}
z^{1}=z^{3}= \frac{1+i}{\sqrt{2}} 
\hspace{3mm} , \hspace{3mm} 
z^{2} = i c
\hspace{3mm} , \hspace{3mm} 
\tilde{z}^{1}=\tilde{z}^{3}= i
\hspace{3mm} , \hspace{3mm} 
\tilde{z}^{2}= \tilde{\chi}_{2} + i \frac{c}{\sqrt{2} \left(1-\rho^2\right)} \ ,
\end{equation}
together with
\begin{equation}
\label{AdS4_VEV_STU_2}
\begin{array}{lclclclclc}
e^{2\phi} = \dfrac{\sqrt{2}}{c} 
& , &
\zeta^{0}  = \rho \sin\alpha
& , &
\zeta^1= \rho \cos\alpha
& , &
\zeta^2 = l \sin\alpha 
& , &
\zeta^3 = -\rho \cos\alpha & , \\[2mm]
\sigma=\sigma
& , &
\tilde{\zeta}_0  = l \sin\alpha
& , &
\tilde{\zeta}_1 = l \cos\alpha
& , &
\tilde{\zeta}_2 = - \rho \sin\alpha  
& , &
\tilde{\zeta}_3 = -l \cos\alpha & .
\end{array}
\end{equation}
This solution is spanned by five moduli $(\sigma,\,\tilde{\chi}_2 \, ;\,\rho,\,l,\,\alpha)$. The gauge symmetries are spanned by the vectors 
\begin{equation}
k_1 = \partial_\sigma
\hspace{5mm} , \hspace{5mm} 
k_2 = \partial_{\tilde{\chi}_2} + 2 \, \rho \, \partial_l
\hspace{5mm} , \hspace{5mm} 
k_3-k_4 = 0
\hspace{5mm} , \hspace{5mm} 
k_3+k_4 =  -4 \, (l^2+\rho^2) \, \partial_\alpha \ .
\end{equation}
This allows us to fix $\,(\sigma,\,\tilde{\chi}_2,\,\alpha) = (0,\,0,\,\tfrac{\pi}{2})$. With this gauge fixing, and a final redefinition of the scalar fields of the form
\begin{equation}
l = \pm \, c \chi 
\hspace{10mm} \text{ and } \hspace{10mm} 
\rho^2 = \frac{\varphi^2}{1+\varphi^2} \ ,
\end{equation}
one finds a mapping between our moduli fields and the ones in \cite{Bobev:2021yya}. Within the STU-model, the spectrum of scalar fluctuations around these AdS$_{4}$ vacua shows a dependence on the two moduli fields $(\chi,\varphi)$\footnote{When further truncating to the S$^2$T-model, the mass spectrum turns out to be independent of $\,\chi\,$.}.

The two moduli fields $(\varphi,\chi)$ parameterise flat directions of the scalar potential which, via the AdS$_{4}$/CFT$_{3}$ correspondence, are dual to marginal operators spanning a conformal manifold of $\mathcal{N}=2$ S-fold CFT$_{3}$'s \cite{Bobev:2021yya}. At generic values of $(\varphi,\chi)$ the corresponding $\textrm{AdS}_{4}$ vacuum preserves $\mathcal{N}=2$ supersymmetry and $\textrm{U}(1)_{1} \times \textrm{U}(1)_{2}$ residual symmetry both within the STU-model and also in the maximal theory. However there are two special cases:
\begin{itemize}

\item \textbf{Case} $\boldsymbol{(\varphi,\chi)=(1,0)}$: this AdS$_{4}$ vacuum features $\mathcal{N}=4$ supersymmetry and an $\textrm{SO}(4)$ residual symmetry enhancement in the maximal theory. The corresponding $\mathcal{N}=4 \,\&\,\textrm{SO}(4)$ S-fold solution of type IIB supergravity was originally presented in \cite{Inverso:2016eet} and later on re-written in a simpler form in \cite{Giambrone:2021wsm}.

\item \textbf{Case} $\boldsymbol{(\varphi,\chi)=(0,0)}$: this AdS$_{4}$ vacuum features $\mathcal{N}=2$ supersymmetry and a $\textrm{U}(2)$ residual symmetry enhancement in the maximal theory. Its uplift to an $\mathcal{N}=2 \,\&\,\textrm{U}(2)$ S-fold solution of type IIB supergravity was presented in \cite{Guarino:2020gfe}. Note that all the scalars $(\zeta^{A},\tilde{\zeta}_{A})$ in (\ref{AdS4_VEV_STU_2}) spanning the Heisenberg fiber of the QK geometry (\ref{ds_QK_STU}) are zero in this solution.

\end{itemize}

\noindent We will focus on these two special AdS$_{4}$ S-fold solutions later on when presenting the type IIB uplift of the universal AdS$_4$ black hole that asymptotes S-fold solutions at infinity.

\subsection{\texorpdfstring{$\textrm{AdS}_{2} \times \Sigma_{\mathfrak{g}}$}{AdS2 x Sigma g} solutions}

Let us now consider non-maximally symmetric solutions. In particular, let us investigate solutions of the form $\textrm{AdS}_{2} \times \Sigma_{\mathfrak{g}}$ with a spacetime metric given by
\begin{equation}
\label{metric_AdS2xSigma2_STU}
d s^2 = - \frac{r^2}{L^2_{\textrm{AdS}_{2}}} \, d t^2 + \frac{L^2_{\textrm{AdS}_{2}}}{r^2} \, d r^2 + L_{\Sigma_{\mathfrak{g}}}^{2} \, d\Omega_{\Sigma_{\mathfrak{g}}} \ ,
\end{equation}
where $\,L_{\textrm{AdS}_{2}}\,$ and $\,L_{\Sigma_{\mathfrak{g}}}\,$ are the AdS$_{2}$ and the $\,\Sigma_{\mathfrak{g}}\,$ radii, respectively. The ansatz for the vector and tensor fields supporting the geometry is given by
\begin{equation}
\label{vector-tensor_STU}
\begin{array}{llll}
\mathcal{A}^\Lambda &=& \mathcal{A}_t{}^\Lambda(r) \, d t - p^\Lambda \, \frac{ \cos \sqrt{\kappa} \, \theta }{ \kappa } \, d \phi & , \\[2mm]
\tilde{\mathcal{A}}_\Lambda &=& \tilde{\mathcal{A}}_t{}_\Lambda(r) \, d t - e_\Lambda \, \frac{ \cos \sqrt{\kappa} \, \theta }{ \kappa } \, d \phi & , \\[2mm]
\mathcal{B}_{\alpha} &=& b_{\alpha}(r) \, \frac{ \sin \sqrt{\kappa} \, \theta }{ \sqrt{\kappa} } \, d \theta \wedge d \phi & ,
\end{array}
\end{equation}
where $\,\kappa= +1\,$ ($\,\kappa= -1\,$) for a spherical (hyperbolic) geometry $\,\Sigma_{\mathfrak{g}}$. Lastly, the scalars $\,z^{i}\,$ and $\,q^{u}\,$ are taken to be constant functions.

It will prove convenient to introduce a vector of charges $\,\mathcal{Q}\,$ of the form 
\begin{equation}
\mathcal{Q}^{M} = \left( \mathfrak{p}^{\Lambda}, \mathfrak{e}_{\Lambda} \right)^{T} \ ,
\end{equation}
with
\begin{equation}
\label{charges_STU}
\mathfrak{p}^{\Lambda}=p^{\Lambda} - \frac{1}{2} \, \Theta^{\Lambda \alpha} \, b_{\alpha}
\hspace{8mm} \textrm{ and } \hspace{8mm}
\mathfrak{e}_{\Lambda}=e_{\Lambda} + \frac{1}{2} \, \Theta_{\Lambda}{}^{\alpha} \, b_{\alpha} \ ,
\end{equation}
so that $\,\mathcal{Q}\,$ depends on the constant vector charges $\,(p^{\Lambda},e_{\Lambda})\,$ as well as on the $\,\theta$-$\varphi\,$ components $\,b_{\alpha}(r)\,$ of the tensor fields in (\ref{vector-tensor_STU}).
As in \cite{Klemm:2016wng}, we can choose the temporal gauge $\,\mathcal{A}_{t}{}^{M}=0\,$ which, when combined with the last equation in (\ref{extra_constraints_2}), implies that $\,\mathcal{Q}'=0\,$ in the BPS equations (\ref{BPS_equations}). Therefore, $\,b_{\alpha}\,$ must also be constant functions and the tensor fields can be gauged away by virtue of the additional tensor gauge symmetry given by a one-form gauge parameter \cite{deWit:2007mt}.

The existence of quarter-BPS solutions with the above ansatz requires a set of algebraic equations, known as the attractor equations, given by \cite{Klemm:2016wng}
\begin{equation}
\label{attractor_eqs_STU}
\begin{split}
\mathcal{Q} & =   \kappa \, L_{\Sigma_{\mathfrak{g}}}^{2} \, \Omega \, \mathcal{M} \, \mathcal{Q} ^{x} \, \mathcal{P}^{x} - 4 \, \textrm{Im}(\bar{\mathcal{Z}} \, \mathcal{V}) \ , \\
\dfrac{L_{\Sigma_{\mathfrak{g}}}^{2}}{L_{\textrm{AdS}_{2}}} & =  -2 \, \mathcal{Z} \, e^{-i \beta} \ , \\[2mm]
\left\langle  \mathcal{K}^{u} , \mathcal{V} \right\rangle & =  0 \ ,
\end{split}
\end{equation}
defined in terms of a central charge $\,\mathcal{Z}(z^{i})=\left\langle \mathcal{Q} , \mathcal{V} \right\rangle\,$, the scalar matrix $\,\mathcal{M}(z^{i})\,$ in (\ref{M_scalar_matrix_N2_STU}) and $\,\mathcal{Q}^{x} = \left\langle  \mathcal{P}^{x} , \mathcal{Q} \right\rangle$. The phase $\,\beta\,$ is associated with the complex function
\begin{equation}
\label{W_function_STU}
W=e^{U} (\mathcal{Z} + i \, \kappa \, L_{\Sigma_{\mathfrak{g}}}^2 \, \mathcal{L})= |W| \, e^{i \beta} \ ,
\end{equation}
which depends on the central charge $\,\mathcal{Z}(z^i)\,$ and a superpotential $\,\mathcal{L}(z^{i},q^{u})=\left\langle \mathcal{Q}^{x} \mathcal{P}^{x} , \mathcal{V} \right\rangle$. The attractor equations (\ref{attractor_eqs_STU}) must be supplemented with a charge quantisation condition
\begin{equation}
\label{quant_cond_STU}
\mathcal{Q}^{x} \, \mathcal{Q}^{x} = 1 \ ,
\end{equation}
and a set of compatibility constraints of the form
\begin{equation}
\label{extra_constraints_STU}
\mathcal{H} \, \Omega \, \mathcal{Q} = 0 
\hspace{8mm} \textrm{ and } \hspace{8mm}
\mathcal{H} \, \Omega \, \mathcal{A}_{t} = 0 \ ,
\end{equation}
where $\,\mathcal{H}=(\mathcal{K}^u)^{T} \, h_{uv} \, \mathcal{K}{}^{v}\,$. The second equation in (\ref{extra_constraints_STU}) is automatically satisfied by the temporal gauge fixing condition $\,\mathcal{A}_{t}=0$. We refer the reader to \cite{Klemm:2016wng} for a detailed derivation of the attractor equations (\ref{attractor_eqs_STU})-(\ref{extra_constraints_STU}) and, more generally, for a derivation of the first-order BPS equations collected in Appendix~\ref{sec:app_BPS_equations} from which (\ref{attractor_eqs_STU})-(\ref{extra_constraints_STU}) follow.

In order to solve the attractor equations we apply the following strategy. We first solve the third equation in (\ref{attractor_eqs_STU}), namely, $\left\langle  \mathcal{K}^{u} , \mathcal{V} \right\rangle = 0$ and then complete the solution by solving the remaining equations in (\ref{attractor_eqs_STU})-(\ref{extra_constraints_STU}). The equation $\left\langle  \mathcal{K}^{u} , \mathcal{V} \right\rangle = 0$ implies 
\begin{equation}
\label{KV_equation_1}
\begin{array}{c}
z^1 \, z^{3} = i 
\hspace{5mm} , \hspace{5mm}
z^2 = i c
\hspace{5mm} , \hspace{5mm}
\tilde{z}^1=\tilde{z}^3 = i \ ,
\\[2mm]
\zeta^1 + \zeta^3 = 0
\hspace{4mm} , \hspace{4mm}
\tilde{\zeta}_1 + \tilde{\zeta}_3 =0
\hspace{4mm} , \hspace{4mm}
\zeta^0 + \tilde{\zeta}_2 =0
\hspace{4mm} , \hspace{4mm}
\zeta^2 - \tilde{\zeta}_0=0 \ ,
\end{array}
\end{equation}
together with
\begin{equation}
\label{KV_equation_2}
(z^1 - z^3) \left( (\zeta^A)^2 + (\tilde{\zeta}_A)^{2} \right) = 0 
\hspace{8mm} \textrm{ for } \hspace{8mm} A=0,1,2,3 \ .
\end{equation}
Therefore, there are two different branches of solutions: $i)$ the first one has $z^1 = z^3 = e^{i \tfrac{\pi}{4}}$ with $(\zeta^{A},\tilde{\zeta}_{A})$ being restricted by (\ref{KV_equation_1}). $ii)$ the second one has $\zeta^A = \tilde{\zeta}_A=0$ with $(z^1,z^3)$ being restricted by (\ref{KV_equation_1}). It is also instructive to look at the first equation in (\ref{extra_constraints_STU}) together with the charge quantisation condition in (\ref{quant_cond_STU}). When combined with (\ref{KV_equation_1}) they give the four conditions
\begin{equation}
\label{charges_condition}
\mathfrak{p}^2 + c \,\mathfrak{e}_2 = 0
\hspace{4mm} , \hspace{4mm} 
-\mathfrak{p}^0 + c \, \mathfrak{e}_0 = 0  
\hspace{4mm} , \hspace{4mm} 
g \, (\mathfrak{p}^1 + \mathfrak{p}^3) = \pm 1
\hspace{4mm} , \hspace{4mm} 
(-\mathfrak{p}^1 + \mathfrak{p}^3) \, |\vec{\zeta}|^{2} = 0 \ ,
\end{equation}
which connect with the discussion below (\ref{Embedding_Tensor_STU}). Recalling that the non-compact $\,\mathbb{R}_{1}\,$ and $\,\mathbb{R}_{2}\,$ factors in the gauge group are dyonically gauged by $\,\mathcal{A}^{2}+c\,\tilde{\mathcal{A}}_{2}\,$ and $\,-\mathcal{A}^{0}+c\,\tilde{\mathcal{A}}_{0}\,$, respectively, the first and second conditions in (\ref{charges_condition}) set the non-compact (magnetic) charges to zero. 
On the other hand, the combinations of $\textrm{U}(1)$-generators\footnote{We denote by $\mathfrak{u}(1)$ a generator of $\textrm{U}(1)$.}
\begin{equation}
\label{U(1)_redefinitions}
\mathfrak{u}(1)_{\textrm{R}} \equiv \mathfrak{u}(1)_{2} - \mathfrak{u}(1)_{1}
\hspace{6mm} \textrm{ and } \hspace{6mm}
\mathfrak{u}(1)_{\perp} \equiv\mathfrak{u}(1)_{2}  + \mathfrak{u}(1)_{1} \ ,
\end{equation}
are respectively gauged by 
\begin{equation}
\label{eq:ARandAperp}
\mathcal{A}^{\textrm{R}} \equiv \mathcal{A}^{1}+\mathcal{A}^{3}
\hspace{6mm} \textrm{ and } \hspace{6mm}
\mathcal{A}^{\perp} \equiv -\mathcal{A}^{1}+\mathcal{A}^{3} \ .
\end{equation}
Then the third condition in (\ref{charges_condition}) sets the $\textrm{U}(1)_{\textrm{R}}$ charge of the solution, whereas the fourth condition in (\ref{charges_condition}) forbids a $\textrm{U}(1)_{\perp}$  charge whenever $|\vec{\zeta}|^{2} \neq 0$.

A detailed study of the full set of attractor equations in (\ref{attractor_eqs_STU})-(\ref{extra_constraints_STU}) gives two classes of solutions with $\Sigma_{\mathfrak{g}}$ being (locally) a hyperboloid $\mathbb{H}^{2}$. The first class has the scalars fixed at their values (\ref{AdS4_VEV_STU_1})-(\ref{AdS4_VEV_STU_2}) in the AdS$_{4}$ S-folds, thus generically having $|\vec{\zeta}|^{2} \neq 0$. As we will show, this class of $\textrm{AdS}_{2} \times \mathbb{H}^{2}$ solutions describes the horizon of the universal AdS$_{4}$ black hole that asymptotes to the $(\varphi,\chi)$-family of $\mathcal{N}=2$ S-folds in Section~\ref{sec:general_N=2_AdS4}. The second class of $\textrm{AdS}_{2} \times \mathbb{H}^{2}$ solutions has $|\vec{\zeta}|^{2} = 0$ and comes along with two arbitrary (real) parameters. Upon tuning of the parameters we will show the existence of a regimen in which scale separation between the AdS$_{2}$ and the $\mathbb{H}^{2}$ factors of the geometry occurs.

\subsubsection{Universal \texorpdfstring{$\textrm{AdS}_{4}$}{AdS4} BH that asymptotes to the \texorpdfstring{$(\varphi,\chi)$}{(phi-chi)}-family of \texorpdfstring{$\mathcal{N}=2$}{N=2} S-folds}
\label{subsec:BHAdS4}

The first class of solutions to the attractor equations (\ref{attractor_eqs_STU})-(\ref{extra_constraints_STU}) generically has $|\vec{\zeta}|^{2} \neq 0$. The scalars are set to their values (\ref{AdS4_VEV_STU_1})-(\ref{AdS4_VEV_STU_2}) at the $(\varphi,\chi)$-family of AdS$_{4}$ S-folds. The charges as well as the $\textrm{AdS}_{2}$ and $\Sigma_{\mathfrak{g}}$ radii get also fixed to the values
\begin{equation}
\label{horizon_conf_STU_universal}
\begin{array}{lcll}
\mathfrak{p}^{0} =  \mp \, \dfrac{c}{2\sqrt{2} \, g}
& \hspace{5mm} , &  \hspace{5mm}
\mathfrak{e}_{0} = \mp \, \dfrac{1}{2\sqrt{2} \, g}  & , \\[3mm]
\mathfrak{p}^{1} = \mathfrak{p}^{3} =  \pm \, \dfrac{1}{2 \, g}
& \hspace{5mm} , &  \hspace{5mm}
\mathfrak{e}_{1} = \mathfrak{e}_{3} = 0  & , \\[3mm]
\mathfrak{p}^{2} =  \pm \, \dfrac{c}{2\sqrt{2} \, g}
& \hspace{5mm} , &  \hspace{5mm}
\mathfrak{e}_{2} = \mp \, \dfrac{1}{2\sqrt{2} \, g}  & , \\[4mm]
\kappa \,  L_{\textrm{AdS}_{2}}= - \dfrac{c^{\frac{1}{2}}}{2\,g} 
& \hspace{5mm} , &  \hspace{5mm}
\kappa \, L^2_{\Sigma_{\mathfrak{g}}}= -\dfrac{c}{2\, g^2}  & .

\end{array}
\end{equation}
Then, provided $g, c>0$, it follows that $\kappa=-1$ so that the horizon geometry is (locally) a two-dimensional hyperboloid $\,\mathbb{H}^{2}$. The phase $\,\beta\,$ in (\ref{attractor_eqs_STU}) gets also fixed to $\,\beta=\frac{\pi}{4} \pm \frac{\pi}{2}\,$ where the $\,\pm\,$ sign is correlated with the ones in (\ref{horizon_conf_STU_universal}). Note that the linear combinations of non-compact vectors $\,\mathcal{A}^{2}+c\,\tilde{\mathcal{A}}_{2}\,$ and $\,-\mathcal{A}^{0}+c\,\tilde{\mathcal{A}}_{0}\,$ entering the gauge connection have no magnetic charge, namely, 
\begin{equation}
\mathfrak{p}^{2} + c \, \mathfrak{e}_{2}=0
\hspace{8mm} \textrm{ and } \hspace{8mm}
-\mathfrak{p}^{0} + c \, \mathfrak{e}_{0}=0 \ .
\end{equation}
On the contrary, the electric dual combinations 
\begin{equation}
\label{orthogonal_charges_fixed}
\mathfrak{p}^{2} - c \, \mathfrak{e}_{2}=\pm\frac{c}{\sqrt{2}g}
\hspace{8mm} \textrm{ and } \hspace{8mm}
\mathfrak{p}^{0} + c \, \mathfrak{e}_{0}=\mp\frac{c}{\sqrt{2}g} \ ,
\end{equation}
are \textit{fixed} to non-zero values but they are associated with linear combinations of vectors not entering the gauge connection.

The above class of $\textrm{AdS}_{2} \times \mathbb{H}^{2}$ solutions of the attractor equations can be seen as the near-horizon geometry of the universal $\textrm{AdS}_{4}$ black hole that asymptotes to the $(\varphi,\chi)$-family of $\mathcal{N}=2$ S-folds. The metric of this universal AdS$_{4}$ black hole is given by (\ref{metric_universal_BH_intro})-(\ref{metric_universal_BH_2_intro}) with the hyperbolic horizon being located at $\,r_{H}^2 =\frac{c}{2 g^2}$. The vectors supporting the black hole take the form (\ref{vector-tensor_STU}) where the charges and the horizon data are given in (\ref{horizon_conf_STU_universal}). Finally, the gravitational entropy density computed from the horizon data (\ref{horizon_conf_STU_universal}) reads
\begin{equation}
\label{entropy_universal_STU}
s = \frac{\text{Area}(\Sigma_\mathfrak{g})}{4} = \frac{L^2_{\Sigma_{\mathfrak{g}}} 4 \pi (\mathfrak{g}-1)}{4} =  \frac{c}{g^2} \,\frac{(\mathfrak{g}-1)\pi}{2}\ ,
\end{equation}
which agrees with (\ref{entropy_universal_intro}) provided (\ref{L2_AdS4}) and, consistently, turns out to be independent of the AdS$_{4}$ moduli $\,(\varphi,\chi)\,$ spanning the conformal manifold of $\mathcal{N}=2$ S-fold CFT's. As a check of the results, we have explicitly verified that the above universal $\textrm{AdS}_{4}$ black hole with constant scalars given in (\ref{AdS4_VEV_STU_1})-(\ref{AdS4_VEV_STU_2}) solves the set of first-order BPS equations presented in Appendix~\ref{sec:app_BPS_equations}.

\subsubsection{A two-parameter family of \texorpdfstring{$\textrm{AdS}_{2} \times \mathbb{H}^{2}$}{AdS2 x H2} solutions}
\label{subsubsection:ScaleSeparatedAdS2}

The second class of solutions to the attractor equations (\ref{attractor_eqs_STU})-(\ref{extra_constraints_STU}) fixes the scalars in the Heisenberg fiber of the QK geometry to zero, namely, 
\begin{equation}
\label{Heis_vanish}
\zeta^A = \tilde{\zeta}_A = 0  \ ,
\end{equation}
as well as the $\Sigma_{\mathfrak{g}}$ radius to
\begin{equation}
\label{L_Sigma_second_family}
\kappa L^2_{\Sigma_{\mathfrak{g}}} = -\frac{c}{2 g^2} \ .
\end{equation}
Taking again $\,g,c>0\,$ sets $\,\kappa=-1\,$ and $\,\Sigma_{\mathfrak{g}}\,$ is locally described by a hyperboloid $\,\mathbb{H}^{2}$. However, unlike for the universal $\textrm{AdS}_{4}$ black hole horizon in (\ref{horizon_conf_STU_universal}), the charges are not completely fixed. This time one finds that
\begin{equation}
\label{charges_second_family}
\mathfrak{p}^2 + c \,\mathfrak{e}_2 = 0
\hspace{5mm} , \hspace{5mm} 
-\mathfrak{p}^0 + c\, \mathfrak{e}_0 = 0
\hspace{5mm} , \hspace{5mm}
\mathfrak{p}^1 = \mathfrak{p}^3 = \pm \frac{1}{2g}
\hspace{5mm} , \hspace{5mm}
\mathfrak{e}^1 = \mathfrak{e}^3 = 0 \ ,
\end{equation}
which again implies that the solutions have zero (magnetic) charge under the non-compact vector fields  $\,\mathcal{A}^{2}+c\,\tilde{\mathcal{A}}_{2}\,$ and $\,-\mathcal{A}^{0}+c\,\tilde{\mathcal{A}}_{0}\,$ entering the gauge connection. Nonetheless, the electric dual combinations $\,\mathfrak{p}^{2} - c \, \mathfrak{e}_{2}\,$ and $\,\mathfrak{p}^{0} + c \, \mathfrak{e}_{0}\,$ remain arbitrary in this class of solutions in contrast to what happened in (\ref{orthogonal_charges_fixed}). These two non-zero charges can be used to introduce a single non-zero complex charge
\begin{equation}
\mathbf{p} \equiv  \frac{2g}{c} \left[ (\mathfrak{p}^{2} - c \, \mathfrak{e}_{2} \, )  -  i \, (\mathfrak{p}^{0} + c \, \mathfrak{e}_{0}) \right] \ ,
\end{equation}
in terms of which the second class of solutions is specified by the relations
\begin{equation}
\label{new_horizon_1_p}
z^{1} \, z^{3} =  i
\hspace{3mm} , \hspace{3mm}
z^1 + z^3 = \mathbf{p}
\hspace{3mm} , \hspace{3mm}
z^{2} = i \, c
\hspace{6mm} , \hspace{6mm}
\tilde{z}^1 = \tilde{z}^3 = i
\hspace{3mm} , \hspace{3mm}
\tilde{z}^{2} = \tilde{\chi}_{2} + i\, \dfrac{c}{{\rm Re}\left[ \mathbf{p} \right]} \  ,
\end{equation}
together with
\begin{equation}
\label{new_horizon_2_p}
e^{2\phi}=\frac{{\rm Im}\left[\mathbf{p} \right]}{c}
\hspace{10mm} \textrm{ and } \hspace{10mm}
\kappa\,L_{{\rm AdS}_2}  = - \frac{c^{\frac{1}{2}}}{\sqrt{2} \, g \, |\mathbf{p}|} \left(z^1 \, \bar{z}^{\bar{3}}+ \bar{z}^{\bar{1}} \, z^{3}\right)^{\frac{1}{2}}  \ .
\end{equation}
Three things are worth noticing. First, the conditions $\textrm{Im}[\tilde{z}^{2}]>0$ and $e^{2 \phi} >0$ require $\textrm{Arg}(\mathbf{p}) \in (0,\frac{\pi}{2})$. Second, the relations (\ref{new_horizon_1_p}) and (\ref{new_horizon_2_p}) possess an exchange symmetry $\,z^{1} \leftrightarrow z^{3}$. Third, the horizon data in (\ref{horizon_conf_STU_universal}) for the universal $\textrm{AdS}_{4}$ black hole that asymptotes to the $\mathcal{N}=2\,\&\,\textrm{U}(2)$ S-fold at $(\varphi,\chi)=(0,0)$ is recovered at $\,{\bf{p}}= 2 \, e^{i \frac{\pi}{4}}$.

Finally, from (\ref{L_Sigma_second_family}) and (\ref{new_horizon_2_p}), one observes that it is possible to introduce a hierarchy between the $\,\textrm{AdS}_{2}\,$ and $\,\mathbb{H}^{2}\,$ factors in the spacetime geometry upon tuning of the free (complex) parameter  $\,\mathbf{p}\,$. More concretely,
\begin{equation}
\frac{L_{\textrm{AdS}_{2}}^2}{L^2_{\mathbb{H}^{2}}} = \frac{1}{ |\mathbf{p}|^{2}} \left(z^1 \, \bar{z}^{\bar{3}}+ \bar{z}^{\bar{1}} \, z^{3}\right)  \ .
\end{equation}
Parameterising the solution of the first relation in (\ref{new_horizon_1_p}) as $\,z^{1}=i/z^{3} \equiv e^{\lambda + i \gamma}\,$ requires $\,\gamma \in (0,\frac{\pi}{2})\,$ for both $\,\textrm{Im}z^{1} > 0\,$ and $\,\textrm{Im}z^{3} > 0\,$. Then one finds that
\begin{equation}
\label{scale-separation_4D}
\frac{L_{\textrm{AdS}_{2}}^2}{L^2_{\mathbb{H}^{2}}} = \frac{1}{1+x}
\hspace{8mm} \textrm{ with } \hspace{8mm}
x=\frac{\cosh(2\lambda)}{\sin(2\gamma)} \ ,
\end{equation}
and a hierarchy $\,L_{\textrm{AdS}_{2}}/L_{\mathbb{H}^{2}}\rightarrow 0\,$ is achieved whenever $\,\lambda \rightarrow \pm \infty\,$ or $\,\gamma = \frac{\pi}{4} \pm \frac{\pi}{4}$. However, having scale separation between AdS$_2$ and $\mathbb{H}^{2}$ in the four-dimensional solution by no means implies that scale separation also occurs in the corresponding ten-dimensional background. We will address this question in the next section when investigating the type IIB uplift of this second class of solutions.

\section{Type IIB uplift}
\label{sec:IIB_uplift}

With the advent of Exceptional Field Theory (ExFT) \cite{Hohm:2013pua} and Generalised Geometry \cite{Coimbra:2011ky,Coimbra:2012af}, a systematic procedure has been established to uplift  maximal gauged supergravities (with a higher-dimensional origin) to string/M-theory \cite{Lee:2014mla,Hohm:2014qga,Inverso:2017lrz}. The procedure, known as generalised Scherk--Schwarz (SS) reduction, is a generalisation of the ordinary SS twisted reduction of \cite{Scherk:1979zr} that uses the exceptional $\,\textrm{E}_{d(d)}\,$ symmetry, with $d=11-D$, of the (ungauged) maximal supergravity in $D$ dimensions as a guiding principle.

The type IIB uplift of the $\,D=4\,$ maximal $\,\left[\textrm{SO}(1,1)\times \textrm{SO}(6)\right]\ltimes \mathbb{R}^{12}\,$ gauged supergravity is encoded in an $\textrm{SL}(8)$ generalised twist matrix $\,U_{\mathcal{M}}{}^{\mathcal{N}}(Y)\,$ and an $\mathbb{R}^{+}$ scaling function $\,\rho(Y)\,$. They depend on the six coordinates $\,y^{m}\,$ of the internal space which are in turn embedded in a larger $56$-dimensional generalised geometry with coordinates $Y^{\mathcal{M}}$ transforming in the fundamental representation of the $\textrm{E}_{7(7)}$. The explicit form of $\,U_{\mathcal{M}}{}^{\mathcal{N}}(Y)\,$ and $\,\rho(Y)\,$ was given in \cite{Inverso:2016eet}. Using this data, the field content of the maximal $D=4$ supergravity (r.h.s of \eqref{gSS_ansatz})\footnote{The field content of the STU-model in Section~\ref{sec:STU-model} is embedded into maximal $D=4$ supergravity as follows. The scalar-dependent matrix $\,\mathcal{M}_{\mathcal{MN}}(x)\,$ is given in (\ref{M_MN_4D}). The vectors $\,\mathcal{A}_{\mu}{}^{\mathcal{M}}(x)\,$ are identified in (\ref{A^M_4D}). Lastly, we do not need to consider the two-form $\mathcal{B}_{1}$ and $\mathcal{B}_{2}$ as they can be gauge fixed to zero (see discussion below (\ref{charges_STU})). Consequently, the scalar currents sourcing them, see \textit{e.g.} (\ref{dB1_current}), do vanish for the specific scalar VEV's and vector charges in the solutions we will uplift.} and the one the $\textrm{E}_{7(7)}$-ExFT \cite{Hohm:2013uia} (l.h.s of \eqref{gSS_ansatz}) are related by a generalised SS ansatz of the form
\begin{equation}
\begin{array}{rcl}
\label{gSS_ansatz}
g_{\mu\nu}(x,Y) &=& \rho^{-2}(Y)\, g_{\mu\nu}(x) \ , \\[2mm]
M_{\mathcal{MN}} (x,Y) &=& U_{\mathcal{M}}{}^{\mathcal{K}}(Y) \, U_{\mathcal{N}}{}^{\mathcal{L}}(Y)\, \mathcal{M}_{\mathcal{KL}}(x) \ ,  \\[2mm]
A_{\mu}{}^{\mathcal{M}}(x,Y) &=& \rho^{-1} \,(U^{-1})_{\mathcal{N}}{}^{\mathcal{M}}(Y) \,\mathcal{A}\indices{_\mu^{\mathcal{N}}}(x) \ ,  \\[2mm]
B_{\mu\nu\,\alpha}(x,Y) &=& \rho^{-2}(Y)\, U\indices{_\alpha^\beta}(Y)\, \mathcal{B}_{\mu\nu\,\beta}(x) \ ,  \\[2mm]
B_{\mu\nu\,\mathcal{M}}(x,Y) &=& -2\,\rho^{-2}(Y)\, (U^{-1})_{\mathcal{S}}{}^{\mathcal{P}}(Y)\,\partial_{\mathcal{M}}U_{\mathcal{P}}{}^{\mathcal{R}}(Y)\, \mathcal{B}_{\mu\nu\,\alpha}(x)\, (t^\alpha)_{\mathcal{R}}{}^{\mathcal{S}} \ .
\end{array}    
\end{equation}
The last step in the uplift procedure requires to use the dictionary between the fields of $\textrm{E}_{7(7)}$-ExFT and those of ten-dimensional type IIB supergravity put forward in \cite{Baguet:2015sma}.\footnote{In this section we set $\,g=c=1\,$ without loss of generality. One can verify that the type IIB uplift following from the generalised SS ansatz in (\ref{gSS_ansatz}) is indeed insensitive to these parameters.}

Employing the above procedure, the maximal $\,\left[\textrm{SO}(1,1)\times \textrm{SO}(6)\right]\ltimes \mathbb{R}^{12}\,$ supergravity in 4D has been shown to describe the dimensional reduction of type IIB supergravity on $\,\textrm{S}^{1} \times \textrm{S}^{5}$. Importantly, the reduction turns out to be \textit{non-geometric}: it incorporates a non-trivial S-duality twist of the type IIB fields when looping along the $\,\textrm{S}^{1}\,$ specified by an $\,\textrm{SO}(1,1) \subset \textrm{SL}(2,\mathbb{R})\,$ twist matrix\footnote{We hope not to create confusion between the S-duality SL(2) fundamental index $\alpha=1,2$ in (\ref{eq:defASL2}), the coordinate $\,\alpha \in [0,\,\frac{\pi}{2}]\,$ along the $\,\textrm{S}^{5}\,$ in (\ref{internal_metric_N4}), and the $\textrm{E}_{7(7)}$ adjoint index $\,\alpha=1,\ldots,133\,$ in (\ref{gSS_ansatz}).}
\begin{equation}
\label{eq:defASL2}
A^\alpha{}_\beta(\eta) = \begin{pmatrix} e^{-\eta} & 0\\ 0 & e^{\eta} \end{pmatrix} \ .
\end{equation}
As a consequence of the S-duality twist, the entire dependence of the type IIB axion-dilaton matrix $\,m_{\alpha\beta}\,$ and two-form potentials $\,\mathbb{B}^{\alpha}\,$ on the coordinate $\,\eta \in [0,T]\,$ along the $\,\textrm{S}^{1}\,$ is through the twist matrix (\ref{eq:defASL2}), namely,
\begin{equation}
\label{m&B_twist}
m_{\alpha\beta} = (A^{-t})_{\alpha}{}^{\gamma} \, \mathfrak{m}_{\gamma\delta}  (A^{-1})^{\delta}{}_{\beta} =
\left(  \begin{array}{cc}
    e^{-\Phi} + e^{\Phi} C_{0}^{2}  & - e^{\Phi} C_{0} \\
    - e^{\Phi} C_{0}  & e^{\Phi}
\end{array}\right)
\hspace{6mm} \textrm{ and } \hspace{6mm}
\mathbb{B}^\alpha = A^\alpha{}_\beta \, \mathfrak{b}^\beta \ ,
\end{equation}
with $\,\mathfrak{m}_{\gamma\delta} \,$ and $\,\mathfrak{b}^\beta\,$ being independent of $\,\eta$. The S-duality twist in (\ref{eq:defASL2}) induces a non-trivial hyperbolic monodromy
\begin{equation}
\label{monodromy}
\mathfrak{M}_{\textrm{S}^{1}} = A^{-1}(\eta) \, A(\eta + T) =  \begin{pmatrix} e^{-T} & 0\\ 0 & e^{T} \end{pmatrix} \ ,
\end{equation}
that can be brought into a generic $\,\textrm{SL}(2,\,\mathbb{Z})\,$ hyperbolic monodromy of the form
\begin{equation}
\label{Jk_monodromy}
J_k = 
\begin{pmatrix}
    k & 1\\
    -1 & 0 
\end{pmatrix} = - \mathcal{S} \, \mathcal{T}^k 
\hspace{8mm} \textrm{with} \hspace{8mm} k \in \mathbb{N} 
\hspace{8mm} \textrm{and} \hspace{8mm}
k > 2 \ ,
\end{equation}
provided the period $\,T(k)\,$ becomes $k$-dependent \cite{Inverso:2016eet}. This renders both the S-folds and the solutions presented in this work full-fledged solutions in type IIB string theory.

In this section we will present the type IIB uplift of the universal black hole of Section~\ref{subsec:BHAdS4} for two particular asymptotics: the $\,\mathcal{N}=2\,$ S-fold with $\,\textrm{U}(2)\,$ symmetry at $\,(\varphi,\chi)=(0,0)\,$ and the $\,\mathcal{N}=4\,$ S-fold with $\,\textrm{SO}(4)\,$ symmetry at $\,(\varphi,\chi)=(1,0)$. The most general universal BH asymptoting the $(\varphi,\chi)$-family of AdS$_{4}$ solutions can be straightforwardly uplifted using the same procedure. However, in the absence of a simple uplift even for the AdS$_4$ solutions at generic values of $\,\varphi$, we will refrain from presenting the lengthy output in this work. Finally, we will also present the type IIB uplift of the $\,\textrm{AdS}_2 \times \mathbb{H}^2\,$ solutions in Section~\ref{subsubsection:ScaleSeparatedAdS2} particularised to the case $\,\gamma = \frac{\pi}{4}\,$. We will show that these solutions uplift to $\,\textrm{AdS}_{2} \times \textrm{M}_{8}\,$ supersymmetric S-fold backgrounds with $\,\textrm{M}_{8}=\mathbb{H}^{2} \times \textrm{S}^{5}  \times \textrm{S}^{1}\,$ that admit parametrically-controlled scale separation, and discuss how the supergravity approximation breaks down in the limit of infinite scale separation.

\subsection{Uplift of the universal BH that asymptotes to the \texorpdfstring{$\mathcal{N}=4\,\&\,\textrm{SO}(4)$}{N=4 SO(4)} S-fold}

We use the conventions and coordinates of \cite{Guarino:2022tlw} to describe the $\textrm{S}^1 \times \textrm{S}^5$ internal geometry. The $\,\textrm{S}^{1}\,$ is parameterised by a periodic coordinate $\,\eta \in [0,\,T]$ of period $\,T\,$ whereas the $\textrm{S}^{5}$ is understood as two 2-spheres $\,\textrm{S}_{1}^{2}\,$ and $\,\textrm{S}_{2}^{2}\,$ with polar and azimuthal angles $(\theta_i,\,\varphi_i)$, with $i=1,\,2$, fibered over an interval $\alpha \in [0,\,\frac{\pi}{2}]$.

The ten-dimensional type IIB metric receives a contribution from the four-dimensional vector $\,\mathcal{A}^{\textrm{R}} \equiv \mathcal{A}^{1}+\mathcal{A}^{3}\,$ in \eqref{eq:ARandAperp} associated with the R-symmetry $\,\mathfrak{u}(1)_{\textrm{R}} \equiv \mathfrak{u}(1)_{2}-\mathfrak{u}(1)_{1}\,$ in (\ref{U(1)_redefinitions}). This vector then becomes the Kaluza--Klein (KK) vector in the dimensional reduction, namely,
\begin{equation}
A_{\mu}^{\textrm{KK}} \equiv {A_\mu}^n \otimes \partial_{n}   = -\frac{1}{2} \cosh \theta\,d\phi \otimes (\partial_{\varphi_2}-\partial_{\varphi_1}) \ ,
\end{equation}
where we have substituted the value of $\,\mathfrak{p}^{1}+\mathfrak{p}^{3}\,$ at the universal BH solution (\ref{horizon_conf_STU_universal}). One then sees that the R-symmetry $\,\textrm{U}(1)_{\textrm{R}}\,$ is geometrically realised as the Killing vector $\partial_{\varphi_2}-\partial_{\varphi_1}$ in $\,\textrm{S}^{5}$. The ten-dimensional metric then reads
\begin{equation}
\label{10D_metric_BH_N4}
ds_{10}^{2} =   \Delta^{-1} \left(\tfrac{1}{2} \, ds^2_{4} + g_{mn} \,  Dy^m \, Dy^n \right)
\end{equation}
where the external spacetime metric $\,ds^2_4\,$ is that of the universal BH in (\ref{metric_universal_BH_intro})-(\ref{metric_universal_BH_2_intro}) and where 
\begin{equation}
\label{eq:CovDerCoord}
Dy^n = dy^n + {A_\mu}^n \, dx^\mu \ .
\end{equation}
The metric $\,g_{mn}\,$ on the internal space $\,\textrm{S}^{1} \times \textrm{S}^{5}\,$ is given by
\begin{equation}
\label{internal_metric_N4}
g_{mn} \, dy^m dy^n = d\eta^2  + d\alpha^2  + \dfrac{\cos^2\alpha}{2+\cos(2\alpha)} ds_{\textrm{S}^2_1}^2  + \dfrac{\sin^2\alpha}{2-\cos(2\alpha)} ds_{\textrm{S}^2_2}^2 \ ,
\end{equation}
with
\begin{equation}
ds_{\textrm{S}^2_i}^2  = d\theta_{i}^2 + \sin^2\theta_{i} \, {d\varphi_{i}}^2\,,
\end{equation}
and the non-singular warping factor reads
\begin{equation}
\Delta^{-4} = 4 - \cos^2(2\alpha) \ . 
\end{equation}

As for the type IIB S-fold backgrounds, the dependence on the coordinate $\eta$ of the type IIB fields transforming under S-duality is encoded in the SL$(2)$ twist matrix in (\ref{eq:defASL2}). The type IIB axion-dilaton matrix is given by
\begin{equation}
\label{eq:defUntwistIIB}
m_{\alpha\beta} = (A^{-t})_{\alpha}{}^{\gamma} \, \mathfrak{m}_{\gamma\delta}  (A^{-1})^{\delta}{}_{\beta} \ ,
\end{equation}
in terms of the $\eta$-independent matrix
\begin{equation}
\mathfrak{m}_{\alpha\beta} = 
\begin{pmatrix} \sqrt{\frac{2 + \cos(2\alpha)}{2-\cos(2\alpha)}} & 0 \\
0 & 
\sqrt{\frac{2 - \cos(2\alpha)}{2 + \cos(2\alpha)}}
\end{pmatrix}  \ .
\end{equation}
The twisted two-form gauge potentials $\,\mathbb{B}^\alpha = A^\alpha{}_\beta \, \mathfrak{b}^\beta\,$ get contributions from the scalars, as well as from the KK and non-compact vectors in 4D. They are given by
\begin{equation}
\begin{array}{rcl}
\mathbb{B}^1 &=& \frac{1}{\sqrt{2}}\frac{1}{r} dt \wedge d(e^{-\eta} \sin\alpha\,\cos\theta_2)+ \frac{\cosh\theta}{2\sqrt{2}} d\phi\wedge d(e^{-\eta} \cos\alpha\,\cos\theta_1) - 2 \sqrt{2}\,e^{-\eta} \, \cos\alpha \, \widetilde{\textrm{vol}}_{1} \ ,\\[4mm]
\mathbb{B}^2 &=& -\frac{1}{\sqrt{2}}\frac{1}{r}  dt \wedge d(e^{\eta} \cos\alpha\,\cos\theta_1) - \frac{\cosh\theta}{2\sqrt{2}} d\phi \wedge d(e^{\eta} \sin\alpha\,\cos\theta_2)  - 2 \sqrt{2}\, e^{\eta}\, \sin\alpha \, \widetilde{\textrm{vol}}_{2} \ ,
\end{array}    
\end{equation}
with
\begin{equation}
\widetilde{\textrm{vol}}_1 = \frac{\cos^2\alpha}{2+\cos(2\alpha)}\sin\theta_1\, d\theta_1 \wedge D\varphi_1
\hspace{5mm} \text{ and } \hspace{5mm} 
\widetilde{\textrm{vol}}_2 = \frac{\sin^2\alpha}{2-\cos(2\alpha)}\sin \theta_2\, d\theta_2 \wedge D\varphi_2  \ .
\end{equation}

The self-dual five-form flux of type IIB is $\eta$-independent and only gets contributions from the KK vector and the scalars. It reads
\begin{equation}
\begin{array}{rcl}
\label{10D_F5_BH_N4}
\widetilde{F}_5 &=&\, 6\, \widetilde{\textrm{vol}}_5 - 4\sin(2\alpha) \, d\eta \wedge \widetilde{\text{vol}}_{1}\wedge\widetilde{\text{vol}}_{2}\\[2mm]
&+& \frac{r^2}{4} \sinh \theta \left(2\,d(\cos^2\alpha)-3 d\eta\right)\wedge dt\wedge dr \wedge d\theta \wedge d\phi\\[2mm]
&+&\frac{1}{r^2}dt\wedge dr \wedge (\cos\theta_2 \widetilde{\text{vol}}_2 - \cos\theta_1 \widetilde{\text{vol}}_1) \wedge \left(d(\cos^2\alpha)-d\eta\right)\\[2mm]
&-&\frac{1}{2\, r^2}dt\wedge dr \wedge \left(\sin^2\theta_1 \, D\varphi_1 + \sin^2\theta_2 \, D\varphi_2\right) \wedge d(\cos^2\alpha)\wedge d\eta\\[2mm]
&-& \frac{\sinh\theta}{2} d\theta \wedge d\phi\wedge \left(\cos\theta_1 \widetilde{\text{vol}}_2 -\cos\theta_2 \widetilde{\text{vol}}_1 \right)\wedge (d\alpha - \sin(2\alpha) d\eta)\\[2mm]
&-& \frac{\sinh\theta}{4}\, \sin(2\alpha)\, d\theta \wedge d\phi\wedge \left(\sin\theta_1 \, d\theta_1 \wedge \widetilde{\text{vol}}_2 + \sin\theta_2 \, d\theta_2 \wedge \widetilde{\text{vol}}_1 \right) \ ,
\end{array}
\end{equation}
where $\,\widetilde{\textrm{vol}}_5 = d\alpha \wedge \widetilde{\textrm{vol}}_{1} \wedge \widetilde{\textrm{vol}}_{2}$. The first line is the Hodge dual of the second one, and the third and fourth lines are the Hodge dual of the fifth and sixth ones. The solution preserves a $\textrm{U}(1)_{\varphi_{1}} \times \textrm{U}(1)_{\varphi_{2}}$ symmetry associated with shifts of the azimuthal angles on the two $2$-spheres. Lastly, we have verified that the uplift presented here solves the equations of motion of type IIB supergravity in the Einstein's frame.

\subsection{Uplift of the universal BH that asymptotes to the \texorpdfstring{$\mathcal{N}=2\,\&\,\textrm{U}(2)$}{N=2 U(2)} S-fold}

We use once again the conventions and coordinates of \cite{Guarino:2022tlw}. The $\,\textrm{S}^1\,$ continues being parameterised by $\,\eta \in [0,\,T]\,$ whereas the $\textrm{S}^5$ is now understood as a three-sphere $\,\textrm{S}^3\,$ fibered over a two-sphere $\,\textrm{S}^2$. The coordinates on $\,\textrm{S}^3\,$ are the three angles $\,\alpha \in [0,\,2\pi]$, $\,\beta \in [0,\,\pi]\,$ and $\,\gamma \in[0,\,4\pi]\,$. On the $\,\textrm{S}^2\,$ we use polar and azimuthal angles $\,\uptheta \in [0,\,\pi]\,$ and $\,\upphi \in [0,\,2\pi]$.\footnote{We use ``upper letters" to distinguish the internal angles $\,\uptheta\,$ and $\,\upphi\,$ from the external spacetime coordinates $\,\theta\,$ and $\,\phi\,$ on the hyperboloid $\,\mathbb{H}^2$.} Lastly, in order to make symmetries more manifest, we also introduce a set of $\,\text{SU}(2)\,$ left-invariant forms on $\,\textrm{S}^3\,$ defined as
\begin{equation}
\label{SU2-inv-forms}
\begin{array}{lll}
\sigma_1 &=& \frac{1}{2}\left(- \sin\alpha\, d\beta + \cos \alpha \sin \beta\, d\gamma\right) \ , \\[2mm]
\sigma_2 &=& \frac{1}{2} \left(\cos\alpha\, d\beta +\sin\alpha \sin\beta\, d\gamma\right) \ , \\[2mm]
\sigma_3 &=& \frac{1}{2} \left(d\alpha + \cos\beta\, d\gamma \right) \ .
\end{array}
\end{equation}

As in the previous case, the ten-dimensional type IIB metric receives a contribution from the four-dimensional vector $\,\mathcal{A}^{\textrm{R}} \equiv \mathcal{A}^{1}+\mathcal{A}^{3}\,$ in \eqref{eq:ARandAperp} associated with the R-symmetry $\,\mathfrak{u}(1)_{\textrm{R}} \equiv \mathfrak{u}(1)_{2}-\mathfrak{u}(1)_{1}\,$ in (\ref{U(1)_redefinitions}). This vector becomes again the KK vector upon the identification
\begin{equation}
A_{\mu}^{\textrm{KK}} \equiv {A_\mu}^n \otimes \partial_{n}   = \cosh \theta\,d\phi \otimes \partial_{\alpha} \ ,
\end{equation}
so that the R-symmetry $\,\textrm{U}(1)_{\textrm{R}}\,$ is geometrically realised as the vector $\partial_{\alpha}$ on $\,\textrm{S}^{5}$. Following (\ref{eq:CovDerCoord}), this leads us to introduce a ``covariantised" version of $\sigma_{3}$ in (\ref{SU2-inv-forms}), namely,
\begin{equation}
\tilde{\sigma}_3 = \tfrac{1}{2} (D\alpha + \cos\beta \,d\gamma) \ .
\end{equation}
With the above definitions, the ten-dimensional metric reads
\begin{equation}
ds_{10}^{2}=\Delta^{-1} \left( \tfrac{1}{2} \, ds^2_{4} + g_{mn} \, Dy^m \, Dy^n \right) \ ,
\end{equation}
where the external metric $ds^2_4$ is again the universal BH metric in \eqref{metric_universal_BH_intro}-\eqref{metric_universal_BH_2_intro}. The metric $\,g_{mn}\,$ on the internal $\,\textrm{S}^{1} \times \textrm{S}^{5}\,$ is now given by
\begin{equation}
\label{metric_10D_solu_N2}
g_{mn} \, dy^m dy^n = \tfrac{1}{2} \left( \, d\eta^2 + ds_{\textrm{S}^2}^2  + \cos^2 \uptheta\left[ 8 \, \Delta^4 \, \left( \sigma_1^2 + \sigma_2^2 \right) +  \sigma_3^2 \right] \, \right) \ , 
\end{equation}
with 
\begin{equation}
    \,ds_{\textrm{S}^2}^2 = d \uptheta^2 + \sin^2 \uptheta \, d\upphi^2 \ .
\end{equation}
The non-singular warping factor reads
\begin{equation}
\Delta^{-4} = 6-2\cos(2 \uptheta) \ .
\end{equation}

The type IIB axion-dilaton $\,m_{\alpha\beta}\,$ has the twisted structure in (\ref{m&B_twist}) and it is specified by the $\eta$-independent (untwisted) matrix
\begin{equation}
\mathfrak{m}_{\alpha\beta} =\tfrac{1}{2} \, \Delta^2 \, \begin{pmatrix} 5-\cos(2 \uptheta) + 2 \sin^2 \uptheta \sin(2\upphi) & 2 \sin^2 \uptheta \cos(2\upphi) \\ 2 \sin^2 \uptheta \cos(2\upphi)&   5 - \cos(2 \uptheta) - 2 \sin^2 \uptheta \sin(2\upphi) \end{pmatrix} \ .
\end{equation}
The twisted two-form gauge potentials $\,\mathbb{B}^\alpha = A^\alpha{}_\beta \, \mathfrak{b}^\beta\,$ get again contributions from the scalars and from the KK and non-compact vectors in 4D. They are given by
\begin{equation}
\begin{array}{rcl}
\mathbb{B}^1 &=& \frac{1}{\sqrt{2}}\frac{1}{r} dt\wedge d(e^{-\eta} \,\sin \uptheta\,\cos\upphi)+ \frac{\cosh \theta}{2\sqrt{2}} d\upphi\wedge d(e^{-\eta} \,\sin \uptheta\,\sin\upphi) \\[2mm]
&-& \frac{e^{-\eta}}{\sqrt{2}}\left(8\Delta^4 \cos^2 \uptheta \sin \uptheta \sin\upphi\, \sigma_1\wedge \sigma_2 - \cos \uptheta(\sin\upphi d \uptheta + \cos\uptheta\,\sin\uptheta\,\cos\upphi d\upphi)\wedge \tilde{\sigma}_3  \right)\  , \\[4mm]
\mathbb{B}^2 &=&-\frac{1}{\sqrt{2}}\frac{1}{r}  dt\wedge d(e^{\eta} \sin \uptheta \sin\upphi)- \frac{\cosh \theta}{2\sqrt{2}} d\upphi\wedge d(e^{\eta} \sin \uptheta\cos\upphi) \\[2mm] 
&+& \frac{e^{\eta}}{\sqrt{2}}\left(8\Delta^4 \cos^2 \uptheta \sin \uptheta \cos\upphi \, \sigma_1\wedge \sigma_2 - \cos \uptheta(\cos\upphi \, d\uptheta - \cos\uptheta \sin \uptheta\,\sin\upphi\,d\upphi)\wedge \tilde{\sigma}_3  \right) \ .
\end{array}
\end{equation}

Finally, the self-dual five-form only receives contributions from the scalars and the KK vector in 4D. It takes the form
\begin{align}
\widetilde{F}_5 \,\, &= \,\, -\frac{r^2 \sinh\theta}{8}\,dt\wedge dr \wedge d\theta\wedge d\phi \wedge\left( d(\sin^2\uptheta\,\cos(2\upphi))+6\,d \eta\right)\\
\nonumber& + 4 \cos^3 \uptheta \,\sin\uptheta\,\Delta^4 \,\big(3 d\uptheta \wedge d\upphi+\sin(2\upphi) d\uptheta \wedge d\eta\\
\nonumber&\hspace{4.5cm}+\cos(2\upphi) \cos\uptheta \sin\uptheta d\phi\wedge d\eta\big) \wedge \sigma_1 \wedge \sigma_2 \wedge \tilde{\sigma}_3\\
\nonumber&+ \frac{1}{r^2} dt\wedge dr \Big[2\cos^2 \uptheta \,\Delta^4\, \sigma_1 \wedge \sigma_2\wedge( d(\sin^2\uptheta\,\cos(2\upphi))+2 d\eta) \\
\nonumber&\phantom{+ \frac{1}{r^2} dt\wedge dr \Big[-} +\frac{\sin\uptheta\cos\uptheta}{4} d\uptheta \wedge \left(\sin^2\uptheta\,d(\cos2\upphi)+ 2 d\eta\right)\wedge \tilde{\sigma}_3\Big]\\
\nonumber&- \sinh\theta d\theta \wedge d\phi \Big[2 \sin^2 \uptheta \,\cos^2\uptheta \, \Delta^4\, \left(d\upphi + \sin(2\upphi) d\eta\right) \wedge \sigma_1 \wedge \sigma_2 \\
\nonumber&\hspace{2.7cm}+ \frac{\cos\uptheta \sin \uptheta}{8} \big(\cos(2\upphi) \sin(2\uptheta) d\upphi\wedge d\eta\\
\nonumber&\hspace{5.5cm}+ 2 \sin (2\,\upphi)\, d\uptheta \wedge d\eta + 2 d\uptheta \wedge d\upphi\big)\wedge \tilde{\sigma}_3\Big] \ .
\end{align}
As in the previous case, we have arranged the different terms so that the first and second lines are the Hodge duals of each other. Moreover, the $dt\wedge dr$ contributions of the third and fourth lines are dual to the $d\theta\wedge d\phi$ contributions of the last two lines. The solution preserves an $\textrm{SU}(2) \times \textrm{U}(1)$ symmetry where $\textrm{U}(1)$ is realised as rotations in the $(\sigma_{1},\sigma_{2})$-plane. Lastly, we have also checked explicitly that the above ten-dimensional solution solves the equations of motion of type IIB supergravity in the Einstein's frame.

\subsection{Uplift of the \texorpdfstring{AdS$_2 \times \mathbb{H}^2$}{AdS2 x H2} solutions and scale separation} 

After having uplifted two examples of the universal BH, we present the type IIB uplift of the $\,\textrm{AdS}_2 \times \mathbb{H}^2\,$ solutions in Section~\ref{subsubsection:ScaleSeparatedAdS2}. Our goal is to establish whether or not the scale separation between AdS$_{2}$ and $\mathbb{H}^{2}$ pointed out in (\ref{scale-separation_4D}) actually extends to a scale-separated AdS$_{2}$ vacuum in ten dimensions within the regime of validity of supergravity. To this end it will be sufficient to determine the 10D metric and the axion-dilaton matrix.

A properly scale-separated AdS$_2$ vacuum requires it not to be fibered over the rest of the geometry. In other words, the KK vector should not have components along the AdS$_2$ coordinates $\,(t,r)\,$ yielding a ten-dimensional type IIB background of the form $\,\textrm{AdS}_{2} \times \textrm{M}_{8}\,$ with $\,\textrm{M}_{8}=\mathbb{H}^{2} \times \textrm{S}^{5}  \times \textrm{S}^{1}$. For the solutions in Section~\ref{subsubsection:ScaleSeparatedAdS2}, this occurs whenever $\,\gamma=\frac{\pi}{4}\,$ in (\ref{scale-separation_4D}), namely, along the $\lambda$-family of solutions parameterised by
\begin{equation}
z^{1}=\frac{i}{z^{3}} = e^{\lambda} \, e^{i \frac{\pi}{4}} \ ,
\end{equation}
with $\,\lambda \in \mathbb{R}$. We will focus on the uplift of this $\lambda$-family of $\,\textrm{AdS}_2 \times \mathbb{H}^2\,$ solutions for which (\ref{scale-separation_4D}) simplifies to
\begin{equation}
\label{L_ratios_lambda}
L_{\textrm{AdS}_2}^2=  \frac{L_{\mathbb{H}^{2}}^2 }{2\cosh^2\,\lambda}
\hspace{8mm}\text{ with }\hspace{8mm} 
L_{\mathbb{H}^{2}}^2 =  \frac{c}{2g^2} \ .
\end{equation}
The AdS$_2$ radius is therefore parametrically smaller than the $\mathbb{H}^{2}$ radius by a factor $\,\cosh^{-1}\lambda$. As we will see in a moment, the two limiting values $\,\lambda \rightarrow \pm \infty\,$ are equivalent since they are related by a discrete symmetry exchanging the roles of $z^1$ and $z^3$ in the solution. Also, the near-horizon geometry of the universal BH asymptoting the $\mathcal{N}=2\,\&\,\textrm{U}(2)$ S-fold in the previous section is recovered at $\,\lambda=0$.

Let us investigate the $\,\textrm{AdS}_{2} \times \textrm{M}_{8}\,$ uplift of the $\lambda$-family of solutions in Section~\ref{subsubsection:ScaleSeparatedAdS2} with $\,\gamma=\frac{\pi}{4}$. We start by introducing a set of internal embedding coordinates $\,\mathcal{Y}^{m} \in \mathbb{R}^{6}\,$, with $\sum\, (\mathcal{Y}^{m})^2 = 1$, adapted to the three commuting isometries $\,\partial_{{\varphi}_{i}}\,$ describing a maximal subgroup $\textrm{U}(1)^{3} \subset \textrm{SU}(4)$ inside the isometry group of $\,\textrm{S}^{5}\,$. These embedding coordinates $\mathcal{Y}^{m}$ are parametrically given by
\begin{equation}
\begin{array}{cccccc}
\mathcal{Y}^{1} = r_1 \,\cos\varphi_1   &\hspace{3mm} , &\hspace{3mm} \mathcal{Y}^{3} = r_2 \,\cos\varphi_2 &\hspace{3mm} , &\hspace{3mm} \mathcal{Y}^{5} = r_3 \,\cos\varphi_3 & , \\[2mm]
\mathcal{Y}^{2} = r_1 \,\sin\varphi_1 &\hspace{3mm} , &\hspace{3mm}  \mathcal{Y}^{4} = r_2 \,\sin\varphi_2 &\hspace{3mm} , &\hspace{3mm} \mathcal{Y}^{6} = r_3 \,\sin\varphi_3 & ,
\end{array}
\end{equation}
in terms of coordinates $\,y^{m}=(r_{i},\varphi_{i})$, with $\,i=1,2,3\,$, that have ranges $\,\varphi_{i}\in [0,2\pi]\,$ and  $\,r_i \in [0,\,1]\,$ so that $\,r_1{}^2 + r_2{}^2 + r_3{}^{2} = 1$. The ten-dimensional geometry can be written in terms of two functions
\begin{equation}
\label{f_functions}
f_1 = \cosh\lambda + (r_1^2 -r_2^2) \sinh\lambda 
\hspace{8mm} \textrm{ and } \hspace{8mm}
f_2 = (1+ r_3^2)\cosh\lambda + (r_1^2 - r_2^2)\sinh\lambda  \ ,
\end{equation}
that only depend on the coordinates $\,r_{i}\,$ and the parameter $\,\lambda\,$. Being independent of the three angles $\,\varphi_{i}$, the functions (\ref{f_functions}) specify an internal geometry that features a $\textrm{U}(1)^{3}$ symmetry. 

The ten-dimensional type IIB metric takes a contribution from the KK vector which is identified with
\begin{equation}
A_{\mu}^{\textrm{KK}} \equiv {A_\mu}^n \otimes \partial_{n}   = \tfrac{1}{2} \cosh\theta d\phi \otimes (\partial_{\varphi_2} - \partial_{\varphi_1}) \ .
\end{equation}
This KK vector is independent of the parameter $\,\lambda\,$ so the R-symmetry $\,\textrm{U}(1)_{\textrm{R}}\,$ is geometrically realised as the Killing vector $\partial_{\varphi_2}-\partial_{\varphi_1}$ in any member of the $\lambda$-family of $\,\textrm{AdS}_2 \times \mathbb{H}^2\,$ solutions. The ten-dimensional metric takes the form 
\begin{equation}
\label{metric_lambda}
ds^2 = \Delta^{-1} \left(  ds_{\text{AdS}_2}^2 + ds_{\mathbb{H}^2}^2 + g_{mn} \, Dy^m \,Dy^n\right) \ ,
\end{equation}
where $Dy^n = dy^n + {A_\mu}^n dx^\mu$ and with $ds_{\text{AdS}_2}^2$ and $ ds_{\mathbb{H}^2}^2$ given in (\ref{metric_AdS2xSigma2_STU}) with the radii in (\ref{L_ratios_lambda}). The warping factor entering (\ref{metric_lambda}) reads
\begin{align}
\label{warping_AdS2}
\Delta^{-4}(r_1,\,r_2,\,\lambda) = 4& \cosh^2\lambda\, f_1\,f_2 \ ,
\end{align}
and the metric on $\,\textrm{S}^{1} \times \textrm{S}^{5}\,$ takes the form 
\begin{equation}
\begin{array}{rcl}
\label{internal_S1xS5_metric_lambda}
g_{mn} \, dy^m \, dy^n & = & \dfrac{d\eta^2 }{2\cosh^2\lambda}  +  \dfrac{e^{-\lambda} {dr_1}^2 + e^{\lambda} {dr_2}^2}{f_2} + \dfrac{1}{2 \cosh\lambda} \left( \dfrac{{dr_3}^2}{f_2}+\dfrac{{r_3}^2}{f_1} {d\varphi_3}^2 \right) \\[4mm]
    &+& \dfrac{1}{2\,f_1} \left( e^{-\lambda} \, r_1^2 \, {d\varphi_1}^2 + e^{\lambda} \, r_2^2 \, {d\varphi_2}^2 + \dfrac{r_1^2 r_2^2}{f_2} \left(e^{-\lambda} \, d\varphi_1+ e^\lambda \, d\varphi_2\right)^2\right) \ .
\end{array}
\end{equation}
The uplift of the type IIB axion-dilaton yields
\begin{equation}
\label{axion-dilaton_AdS2}
m_{\alpha\beta} = {\left(R_{\varphi_{3}} A\right)^{-t}} \, \begin{pmatrix} \sqrt{\dfrac{f_1}{f_2}} & 0 \\ 0 & \sqrt{\dfrac{f_2}{f_1}}\end{pmatrix}  {\left(R_{\varphi_{3}} A\right)^{-1}} \ ,
\end{equation}
in terms of the $A$-twist matrix in (\ref{eq:defASL2}) and the rotation matrix along the $X_{5}X_{6}$-plane
\begin{equation}
R_{\varphi_{3}} = \begin{pmatrix}
        \cos\varphi_3 & \sin\varphi_3 \\ 
        -\sin\varphi_3 & \cos\varphi_3 
\end{pmatrix} \ .
\end{equation}
Note the symmetry under the exchange $\,(r_{1},\varphi_{1}) \leftrightarrow (r_{2},\varphi_{2})\,$ followed by $\,\lambda \leftrightarrow -\lambda\,$. This makes it sufficient to explore the range $\,\lambda \geq 0$. 
Finally, from the ten-dimensional metric in (\ref{metric_lambda}) and (\ref{internal_S1xS5_metric_lambda}), we can read off the relevant scales involved in the solution. These are
\begin{equation}
\label{various_scales}
L_{\textrm{AdS}_2} \propto \frac{1}{\cosh\lambda}    
\hspace{5mm} , \hspace{5mm}
L_{\mathbb{H}^2} = \frac{1}{\sqrt{2}}
\hspace{5mm} , \hspace{5mm}
L_{\textrm{S}^1_{\eta}} \propto \frac{T}{\cosh\lambda}
\hspace{5mm} , \hspace{5mm}
\text{vol}_{\textrm{S}^5} \propto \cosh^5\lambda \ ,
\end{equation}
where $\,T\,$ denotes the period of the coordinate $\,\eta\,$ on the internal $\,\textrm{S}_{\eta}^{1}\,$. The period $\,T(k)\,$ is independent of the scaling parameter $\,\lambda\,$ and is determined by the integer $\,k\,$ specifying the $J_{k} \in \mathrm{SL}(2,\,\mathbb{Z})$ monodromy in (\ref{Jk_monodromy}). The apparent shrinking of $L_{\textrm{AdS}_{2}}$ when $\lambda \rightarrow \infty$ can be reinterpreted as a change of units by virtue of the trombone symmetry of type IIB supergravity. We always assume scales to be larger than $\sqrt{\alpha'}$ (\textit{e.g.} $L_{\text{AdS}_2}^2\gg\alpha'$) for the supergravity approximation to be reliable.  As a result, we conclude that the parametrically-controlled scale separation holds in ten dimensions.

\subsubsection*{SUGRA and EFT approximations in the limit $\,\lambda \rightarrow \infty\,$} 

In the limit of large scale separation, \textit{i.e.} $\lambda \rightarrow \infty$, one must check whether the supergravity (SUGRA) approximation is still a valid approximation. Otherwise, various corrections to the solution, \textit{e.g.}, higher-derivative corrections, should be taken into account. The SUGRA regime corresponds to the first order expansion in both $\,g_s\,$ and $\,\alpha' R\,$, where $\,g_{s}\,$ is the string coupling and $\,R\,$ corresponds to any tensor built from the curvature. In other words, we must verify that $g_s \sim e^\Phi \ll 1$ and $\alpha' R \ll 1$. From the axion-dilaton matrix (\ref{axion-dilaton_AdS2}), we observe that the relevant ratio $f_{2}/f_{1}$ determining $\,e^{\Phi}\,$ in the parameterisation of (\ref{m&B_twist}) remains finite in the scale-separated limit $\,\lambda \rightarrow \infty$. More concretely,
\begin{equation}
\left. \frac{f_{2}}{f_{1}} \right|_{\lambda\rightarrow \infty} = \frac{2(1-r_{2}^{2})}{1+r_{1}^{2}-r_{2}^{2}} + \mathcal{O}(e^{-2\lambda}) \ .
\end{equation}
However the curvature corrections become large in the limit of infinite separation $\,\lambda \rightarrow \infty\,$. The ten-dimensional Ricci tensor diverges in this limit signaling a breaking of the SUGRA approximation. Therefore, we cannot trust our solution in this limit.

On the other hand, any two-dimensional effective field theory (EFT) description of this $\,\textrm{AdS}_{2} \times \textrm{M}_{8}\,$ solution should also break down in the limit $\lambda \rightarrow \infty$. From the internal geometry in (\ref{internal_S1xS5_metric_lambda}), we expect a tower of light KK modes to appear in this limit. The reason why is the following. Since KK masses must be computed in units of the AdS$_{2}$ radius $\,L_{\textrm{AdS}_{2}}\,$ in (\ref{various_scales}), the order of magnitude of such KK masses is given by
\begin{equation}
m L_{\textrm{AdS}_{2}} \propto  \dfrac{1}{\ell \cosh\lambda} \ ,
\end{equation}
where $\,\ell\,$ denotes a characteristic length in the internal space. For the $\,\textrm{S}^1_\eta$ circle in the internal geometry we find a regular behaviour
\begin{equation}
m L_{\textrm{AdS}_{2}} \propto \frac{1}{L_{\textrm{S}^{1}_{\eta}} \cosh\lambda} \sim \mathcal{O}({T^{-1}}) \ ,
\end{equation}
when $\,\lambda \rightarrow \infty$. However, the internal $\,\textrm{S}^{5}\,$ becomes singular when $\,\lambda \rightarrow \infty\,$ and we expect a tower of light modes with masses 
\begin{equation}
m L_{\textrm{AdS}_{2}} \propto \frac{1}{L_{\textrm{S}^{5}} \cosh\lambda} \sim \frac{1}{\cosh^2\lambda} \ ,
\end{equation}
coming from the KK modes propagating along the $(r_2,\,\varphi_2)$ directions. This would cause the breaking of a standard two-dimensional EFT description.

\section{Summary and final remarks}
\label{sec:conclusions}

In this work we have embedded the universal AdS$_{4}$ black hole solution of \cite{Romans:1991nq,Caldarelli:1998hg} into type IIB supergravity. We have done it by first constructing such a black hole as a solution of the $\,{\mathcal{N}=2}\,$ STU-model describing a $\,\mathbb{Z}_2 \times \mathbb{Z}_{2}\,$ invariant sector of the $\,[\textrm{SO}(6) \times \textrm{SO}(1,1)]  \ltimes \mathbb{R}^{12}\,$ maximal supergravity in 4D and then, by employing techniques from exceptional field theory (in particular from the $\textrm{E}_{7(7)}$-ExFT), uplifting it to ten-dimensional type IIB supergravity. From a 4D perspective, the universal BH is quarter-BPS (it preserves two real supercharges) and can asymptote \textit{any} of the $(\varphi,\chi)$-family of AdS$_{4}$ solutions dual to the $\,{\mathcal{N}=2}\,$ conformal manifold of S-fold CFT$_{3}$'s \cite{Bobev:2021yya}. However, for the sake of clarity, we have presented the type IIB ten-dimensional uplift only for two particular asymptotics: the $\,\mathcal{N}=2\,$ S-fold with $\,\textrm{U}(2)\,$ symmetry at $\,(\varphi,\chi)=(0,0)\,$ and the $\,\mathcal{N}=4\,$ S-fold with $\,\textrm{SO}(4)\,$ symmetry at $\,(\varphi,\chi)=(1,0)$.

In addition to the universal AdS$_{4}$ black hole, we have also presented a two-parameter $(\lambda,\gamma)$-family of $\,\textrm{AdS}_{2} \times \mathbb{H}^{2}\,$ solutions which are also quarter-BPS within the $\,{\mathcal{N}=2}\,$ STU-model. However, these horizon-like solutions cannot be extended to a full-fledged universal BH, namely, they do not describe the near-horizon geometry of a universal BH. The reason for this is that the scalars are fixed by the attractor equations (\ref{attractor_eqs_STU})-(\ref{extra_constraints_STU}) to constant values that do not extremise the scalar potential. Upon uplift to ten dimensions, it turns out that the choice $\,\gamma=\frac{\pi}{4}\,$ decouples $\,\textrm{AdS}_{2}\,$ from the rest of the geometry and yields a $\lambda$-family of supersymmetric $\,\textrm{AdS}_{2} \times \textrm{M}_{8}\,$ S-fold backgrounds with $\,\textrm{M}_{8}=\mathbb{H}^{2} \times \textrm{S}^{5}  \times \textrm{S}^{1}$. These can be interpreted as AdS$_{2}$ supersymmetric S-folds in type IIB supergravity. To our knowledge, these are the first examples of such solutions. We observe that the parameter $\,\lambda\,$ allows for a parametrically-controlled scale separation between $\,\textrm{AdS}_{2}\,$ and $\,\textrm{M}_{8}\,$ although the limit of infinite separation $\,\lambda \rightarrow  \infty\,$ becomes pathological. The opposite limit $\,\lambda \rightarrow 0\,$ recovers the ten-dimensional near-horizon geometry of the universal BH that asymptotes to the $\,\mathcal{N}=2\,$ S-fold with $\,\textrm{U}(2)\,$ symmetry.

The universal BH that asymptotes to the $\,\mathcal{N}=2\,$ S-fold with $\,\textrm{U}(2)\,$ symmetry plays a central role in the above story as it sits at the intersection between the two classes of solutions we have presented. On the one hand, it is the universal BH that asymptotes to the AdS$_{4}$ vacuum located at $\,(\varphi,\chi)=(0,0)\,$ in the conformal manifold of S-fold CFT$_{3}$'s. On the other hand, its near-horizon region describes the $\,\textrm{AdS}_{2} \times \mathbb{H}^{2}\,$ solution at $\,(\lambda,\gamma)=(0,\frac{\pi}{4})$. This raises the issue of whether non-universal black holes with running scalars exist that connect an $\,\textrm{AdS}_{2} \times \mathbb{H}^{2}\,$ solution with $\,\lambda \neq 0\,$ at the horizon (IR) with an AdS$_{4}$ vacuum with $\,(\varphi,\chi) \neq (0,0)\,$ at infinity (UV). Such non-universal black holes must have running hyperscalars $\,(\zeta^A,\tilde{\zeta}_A\,)$, thus making their analytic study more difficult. Still, a numerical exploration following the strategy in \cite{Guarino:2017eag} could be performed. The BPS equations in the Appendix~\ref{sec:app_BPS_equations} can be solved numerically starting from the desired $\,\textrm{AdS}_{2} \times \mathbb{H}^{2}\,$ horizon solution with $\,\lambda \neq 0\,$ and perturbing it with irrelevant deformations describing how the solutions arrive at the $\,\textrm{AdS}_{2} \times \mathbb{H}^{2}\,$ geometry in the IR. Upon tuning of the deformation parameters, an AdS$_{4}$ black hole (if it exists at all) could be numerically reconstructed. However, we expect the generic AdS$_{2}$ flow constructed in this way to approach the four-dimensional incarnation of the D3-brane solution at $\,r\rightarrow \infty\,$ (UV), in analogy with the AdS$_{3}$ flows constructed in \cite{Guarino:2022tkh} or the Mkw$_{3}$ flows (dual to RG-flows) constructed in \cite{Guarino:2021kyp}. From a holographic perspective, one such generic flows would describe a supersymmetric flow across dimensions which is triggered by the action of a topological twist and that connects (an anisotropic deformation of) $\,\textrm{SYM}_{4}\,$ placed on $\,\Sigma_{\mathfrak{g}} \times \textrm{S}^{1}\,$ in the UV (with an ${\rm SL}(2,\mathbb{Z})$-monodromy acting along $\textrm{S}^1$) to a supersymmetric quantum mechanics in the IR. We leave this dual field theory analysis for future investigation.

On the other hand, the $\lambda$-family of $\,\textrm{AdS}_{2} \times \textrm{M}_{8}\,$ solutions we have presented turned out to accommodate a parametrically-controlled scale separation. It would be interesting to understand in more general terms which gauged $\,\mathcal{N}=2\,$ supergravities admit scale-separated AdS$_2$ solutions as solutions of the attractor equations \eqref{attractor_eqs_STU}. For example, do they require the gauging of non-compact groups or some specific matter content?  It would also be interesting to investigate under what circumstances these scale-separated AdS$_2$ solutions are compatible with the supergravity approximation and, if so, if they admit a two-dimensional effective field theory description. If still applicable in two dimensions, the distance conjecture of \cite{Ooguri_2007} may provide some obstruction to the existence of a SUGRA/2d EFT description of such solutions. In its original formulation, this conjecture predicts a breaking of any EFT description when a scalar field VEV is sent to infinite distance in moduli space. This breaking would be reflected in the appearance of a light tower of KK or winding modes whose mass goes as $m \sim \exp(-\alpha \,\Delta \phi)$. Several generalisations involving parameters which are not scalar VEV's (like black hole charges \cite{Bonnefoy_2020} or cosmological constants \cite{Lust_2019}) require to extend the notion of distance beyond the geodesic distance on the scalar moduli space. Our $\lambda$-family of solutions places us in a similar situation, namely, that of a space of solutions of an EFT, the type IIB supergravity. Since $\,\lambda\,$ is not a modulus of a scalar potential, this analysis calls for an appropriate notion of distance along the lines of \cite{Li:2023gtt,Palti:2024voy}. If such a generalised notion of distance admits points at infinity, it would be interesting to investigate whether and how the EFT/SUGRA approximations break down when approaching those points. One possibility could be to explicitly compute the spectrum of KK excitations from a two-dimensional perspective using KK spectrometry techniques \cite{Malek2020} suitably extended/adapted to the $\textrm{E}_{9}$-ExFT  \cite{Bossard:2018utw,Bossard:2021jix}. We leave this and related issues for (your?) future investigation.

\section*{Acknowledgements}

The work of AG is supported by the Spanish national grant MCIU-22-PID2021-123021NB-I00. The research of AR is supported by NRF-TWAS doctoral Grant MND190415430760, UID 121811 and the Rodrigues J MITP Grant. AR and MT wish to thank Kevin Goldstein for useful discussions. The work of C.S. has been supported by an INFN postdoctoral fellowship, Bando 24736.

\appendix

\section{Systematics of \texorpdfstring{$\,\mathcal{N}=2\,$}{N=2} truncations}
\label{app:N=2_models}

Here we discuss consistent truncations to $\mathcal{N}=2$ models which capture the $\mathcal{N}=4\,\&\,\textrm{SO}(4)$ AdS$_{4}$ vacuum. The couplings of the maximal theory expanded about this solution are encoded in the $T$-tensor evaluated on the corresponding point $\phi_0$ in the moduli space. If $H_0\subset {\rm SU}(8)$ is an invariance of this $T$-tensor, it is a symmetry of the effective theory at $\phi_0$ and therefore, restricting to the singlet sector of $H_0$ defines a consistent truncation featuring $\phi_0$ as a vacuum. Clearly, $H_0$ contains the compact symmetry ${\rm SO}(4)$ of the vacuum. In fact, it also contains a discrete extension thereof.

The $T$-tensor at $\phi_0$ is described by the complex fermion-shift $A_{1\,ij},\,A_{2\,i}{}^{jkl}$ and their complex conjugates, computed in the same point.
We shall consider the ${\rm SU}(8)$ basis in which $A_{1\,ij}$ is diagonal.
We split correspondingly the R-symmetry indices $i,j,\dots$ into $a_1,b_1,\dots=1,\dots, 4$, labelling the broken supersymmetries and $a_2,b_2,\dots=5,\dots, 8$, labelling the four preserved ones. In this basis, the non-vanishing entries of the two tensors read \cite{Gallerati:2014xra}:\footnote{In eqs. \eqref{A1A2}, we use the convention that the indices $a,b,\dots$, with no subscript, on the right-hand sides run from $1$ to $4$ and coincide with $a_1,b_1,\dots$, or with $a_2-4,b_2-4,\dots$, if the same letter on the left-hand side occurs with subscript 1 or 2, respectively.}
\begin{align}\label{A1A2}
    A_{1\, a_1 b_1}\,L_{{\rm AdS}_4}&= \sqrt{2}\,\delta_{a_1 b_1}\,\,,\,\,\,\, A_{1\, a_2 b_2}\,L_{{\rm AdS}_4}= \frac{1}{\sqrt{2}}\,\delta_{a_2 b_2}\,,\nonumber\\
    A_{2\,a_1}{}^{b_1 c_1 d_2}\,L_{{\rm AdS}_4}&=i\,\epsilon^{ab c d }\,\,,\,\,\,\,A_{2\,a_1}{}^{b_1 c_2 d_2}\,L_{{\rm AdS}_4}=-\sqrt{2}\,\delta_{ab}^{c d}\,,
\end{align}
From inspection of the  explicit forms of the two complex tensors,
one finds that the most general element $h$ of ${\rm SU}(8)$ leaving them invariant
has the form:
\begin{equation}
h= \left(\begin{matrix}{M_1} & {\bf 0}\cr {\bf 0} & {M_2}\end{matrix}\right)\,,
\end{equation}
where
\begin{equation}
    M_1=(M_{1\,a_1}{}^{b_1})\in {\rm SO}(4)\,,
\end{equation}
and 
$$M_2=(M_{2\,a_2}{}^{b_2})=\frac{1}{{\rm det}(M_1)}\,M_1\in {\rm O}(4)\,.$$
Therefore, if $M_1\in {\rm SO}(4)$, $M_2=M_1$, while if  $M_1\in {\rm O}(4)$, ${\rm det}(M_1)=-1$ and $M_2=-M_1$.
The group $H_0$ can therefore be a subgroup of ${\rm O}(4)$ and not just of the residual ${\rm SO}(4)$ gauge symmetry. When ${\rm det}(M_1)=-1$, we can write $h$ in the form:
\begin{equation}
\left(\begin{matrix}{\rm SO}(4)\,\mathcal{O} & {\bf 0}\cr {\bf 0} &-{\rm SO}(4)\,\mathcal{O}\end{matrix}\right)\,,
\end{equation}
where
$\mathcal{O}$ is any ${\rm O}(4)$ with determinant $-1$. We can fix, with no loss of generality, $\mathcal{O}={\rm diag}(-1,1,1,1)$.\\
We look for consistent truncations to $\mathcal{N}=2$ supergravities which are singlet sectors of a suitable $H_0$, highlighting the relevant choices for $H_0$.
We start considering $H_0\subset {\rm  SO}(4)$. 
The eight gravitinos transform, with respect to ${\rm SO}(4)$ in the
\begin{equation}
    {\bf 8}=\left({\bf \frac{1}{2}},\,{\bf \frac{1}{2}}\right)\oplus  \left({\bf \frac{1}{2}},\,{\bf \frac{1}{2}}\right)\,.
\end{equation}
If $H_0={\rm SO}(4)$ or $H_0= {\rm U}(2)\subset {\rm SO}(4)$, the decomposition of the supercharges under $H_0$ features no $H_0$-singlet and the truncation is $\mathcal{N}=0$. If $H_0\subset {\rm SO}(3)_d$ diagonal of the two ${\rm SO}(3)$ subgroups of $ {\rm SO}(4)$, then we have at least two $H_0$-singlets among the gravitinos  and the truncation is at least $\mathcal{N}=2$, since, under ${\rm SO}(3)_d$
we have the following branching:
\begin{equation}
    {\bf 8}\rightarrow 2\times {\bf 1}\oplus 2\times {\bf 3}\,.\label{81133}
\end{equation}
If $H_0={\rm SO}(3)_d$ the resulting $\mathcal{N}=2$ truncation has 1 vector and two quaternionic multiplets. The scalar manifold is:
\begin{equation}
\mathcal{M}_{scal}=\frac{{\rm SL}(2,\mathbb{R})}{{\rm SO}(2)}\times \frac{{\rm G}_{2(2)}}{{\rm SU}(2)\times {\rm SU}(2)}\,.
\end{equation}
The complex scalar in the vector multiplet is described by a $t^3$-model. Within this truncation the $\mathcal{N}=4$ vacuum is $\mathcal{N}=1$, since there is just one singlet among the preserved supersymmetries. Choosing $H_0$ to be a non-abelian (discrete) subgroup of ${\rm SO}(3)_d$ may enlarge the $\mathcal{N}=2$ truncation to one with two vector multiplets defining a $st^2$ model and a quaternionic K\"ahler manifold which is still the c-map of the special K\"ahler one, namely ${\rm SO}(4,3)/{\rm SO}(4)\times {\rm SO}(3)$. The $\mathcal{N}=4$ vacuum is still $\mathcal{N}=1$.
We can instead consider $H_0$ to be a dihedral group ${\rm D}_{k}$, of order $2k$, in ${\rm O}(3)_d\subset {\rm O}(4)$, generated by a rotation by $2\pi/k$ about the 3rd axis within the ${\bf 3}$ and a reflection in a plane containing the 3rd axis. The resulting truncation is still the above $(n_v,n_h)=(2,3)$ model, where now the $\mathcal{N}=4$ vacuum is $\mathcal{N}=2$. The two preserved supersymmetries are one of the singlets in \eqref{81133} and one of the directions in one of the two ${\bf 3}$ (3rd axis). If $k=2$, the Dihedral group is isomorphic to $\mathbb{Z}_2\times \mathbb{Z}_2$, generated by the matrices $\mathcal{O}_1,\,\mathcal{O}_2$ in \eqref{Z2xZ2_action}, and the truncation enlarges to the $(n_v,n_h)=(3,4)$ model considered in this paper,  with scalar manifold:
\begin{equation}
\mathcal{M}_{scal}=\left(\frac{{\rm SL}(2,\mathbb{R})}{{\rm SO}(2)}\right)^3\times \frac{{\rm SO}(4,4)}{{\rm SO}(4)\times {\rm SO}(4)}\,.
\end{equation}
The matrix forms of $\mathcal{O}_1,\,\mathcal{O}_2$ in the ${\bf 8}$ of ${\rm SU}(8)$, in the basis in which $A_{1\,ij}$ is diagonal, are:
\begin{align}
    \mathcal{O}_1={\rm diag}(-1,-1,1,1,-1,-1,1,1)\,,\nonumber\\
     \mathcal{O}_2={\rm diag}(1,-1,-1,-1,-1,1,1,1)\,,
\end{align}
From the above expressions, it is apparent that the resulting $\mathbb{Z}_2\times \mathbb{Z}_2$ leaves only the two gravitinos along directions 7,8, invariant. These also belong to the massless supergravity multiplet at the $\mathcal{N}=4$ vacuum.
Therefore, in this model, the $\mathcal{N}=4$ vacuum is $\mathcal{N}=2$. Note that inverting the sign of $\mathcal{O}_2\in {\rm O}(4)$ does not alter the $\mathcal{N}=2$ truncation, whose supersymmetries are, however, embedded in the $\mathcal{N}=8$ ones differently (along directions 3,4). Within this truncation the $\mathcal{N}=4$ vacuum appears as $\mathcal{N}=0$.

\section{First-order BPS equations}
\label{sec:app_BPS_equations}

Within the context of $\,\mathcal{N}=2\,$ supergravity coupled to vector multiplets and hypermultiplets, a set of first-order BPS equations was derived in \cite{Klemm:2016wng} using an ansatz for the spacetime metric of the form
\begin{equation}
\label{metric_universal_appendix}
d s^2 =  - e^{2 U(r)} d t^2 + e^{-2 U(r)} d r^2 + e^{2 (\psi(r) - U(r))}  d\Omega_{\Sigma_{\mathfrak{g}}}  \ , 
\end{equation}
and a vector/tensor ansatz of the form (\ref{vector-tensor_STU}). The set of BPS equations reads
\begin{equation}
\label{BPS_equations}
\begin{split}
U' & =  - e^{-2(\psi-U)} \,  e^{-U} \, \textrm{Re}(e^{-i\beta}\, \mathcal{Z}) - \kappa \, e^{-U} \, \textrm{Im}(e^{-i\beta}\, \mathcal{L}) \ , \\[2mm]
\psi' & =  - 2 \, \kappa \, e^{-U} \, \textrm{Im}(e^{-i\beta}\, \mathcal{L}) \  , \\[2mm]
\mathcal{V}' & =  e^{i\beta} \, e^{-2(\psi-U)} \,  e^{-U} \, (-\frac{1}{2} \, \Omega \, \mathcal{M} \, \mathcal{Q} - \frac{i}{2} \, \mathcal{Q} + \mathcal{Z} \, \bar{\mathcal{V}}  ) \\
& \quad - \, i \, \kappa \, e^{i\beta} \, e^{-U} (-\frac{1}{2} \, \Omega  \, \mathcal{M} \, \mathcal{P}^{x} \mathcal{Q}^{x} - \frac{i}{2} \, \mathcal{P}^{x}\, \mathcal{Q}^{x} + \mathcal{L} \, \bar{\mathcal{V}}  ) - \, i \, A_{r} \, \mathcal{V} \ , \\[2mm]
{q^u}' & =  \kappa \, e^{-U} \, h^{uv} \,  \textrm{Im}(e^{-i\beta}\, \partial_{v} \mathcal{L})  \ , \\[2mm]
\mathcal{Q}' & =  - 4 \, e^{2 (\psi - U)} e^{-U}  \mathcal{H} \, \Omega \, \textrm{Re}(e^{-i\beta}\, \mathcal{V}) \ , \\[2mm]
\mathcal{\beta}' & =   2 \, \kappa \, e^{-U} \, \textrm{Re}(e^{-i\beta}\, \mathcal{L}) - A_{r}  \ ,
\end{split}
\end{equation}
where the prime denotes a radial derivative and $\,A_{r} = \textrm{Im}\left[  (z^{i})'\partial_{z^{i}}K \right]\,$ is the U(1) K\"ahler connection in $\,\mathcal{M}_{\textrm{SK}}\,$. The  system \eqref{BPS_equations} must be supplemented with the charge quantisation condition in (\ref{quant_cond_STU}) and a set of additional constraints
\begin{equation}
\label{extra_constraints_2}
\mathcal{H} \, \Omega \, \mathcal{Q} = 0 \ , \qquad
h_{uv} \, \mathcal{K}_{M}{}^{u} \, {q^{v}}' = 0 \ , \qquad
\mathcal{H} \, \Omega \, \mathcal{A}_{t} = 2 \, e^{U} \, \mathcal{H} \, \Omega \, \textrm{Re}(e^{-i\beta} \mathcal{V}) \ .
\end{equation}
We refer to \cite{Klemm:2016wng} for more details about the derivation of the BPS equations and the additional constraints.

\bibliography{references}

\end{document}